\documentclass[journal]{IEEEtran}
\usepackage{algorithm}
\usepackage{epsfig}
\usepackage{subfigure}
\usepackage{multirow}
\usepackage{float}
\usepackage{graphicx}
\usepackage[centerlast]{caption2}
\usepackage{makecell,rotating}
\usepackage{booktabs}
\usepackage{array}
\usepackage{stfloats}
\usepackage{amsmath}
\usepackage{url}
\usepackage{bigstrut}
\usepackage{color}
\usepackage{indentfirst}
\usepackage{algorithmicx}
\usepackage{algpseudocode}
\usepackage{verbatim}
\usepackage{graphicx}
\begin{document}
\title{A Generalized Rate-Distortion-$\lambda$ Model Based HEVC Rate Control Algorithm}

\author{Minhao Tang$^{1}$, Jiangtao Wen$^{1}$, Yuxing Han$^{2}$
{\small\begin{minipage}{\linewidth}\begin{center}
\begin{tabular}{ccc}
$^{1}$Tsinghua University, Beijing, China & \hspace*{0.5in} & $^{2}$South China Agricultural University, Guangzhou, China \\
\url{jtwen@tsinghua.edu.cn} && \url{yuxinghan@scau.edu.cn}
\end{tabular}
\end{center}\end{minipage}}
\thanks{This work is supported by Boyan Information 
Technology Ltd and Natural Science Foundation of China 
(Project Number 61521002).}
}

\maketitle

\begin{abstract}
The High Efficiency Video Coding (HEVC/H.265) standard 
doubles the compression efficiency of the 
widely used H.264/AVC standard. For practical applications, rate control 
(RC) algorithms for HEVC need to be developed. Based on the 
R-Q, R-$\rho$ or R-$\lambda$ models, rate control algorithms 
aim at encoding a video clip/segment to a target bit rate 
accurately with high video quality after compression.
Among the various models used by HEVC rate control algorithms, the 
R-$\lambda$ model performs the best in both coding efficiency and rate control accuracy. However,
compared with encoding with a fixed quantization parameter (QP),
even the best rate control algorithm \cite{li2018lambda} still under-performs when comparing the 
video quality achieved at identical average bit rates. 

In this paper, we propose a novel generalized rate-distortion-$\lambda$ (R-D-$\lambda$)
model for the relationship between rate (R), 
distortion (D) and the Lagrangian multiplier ($\lambda$) in rate-distortion (RD) optimized 
encoding. In addition to the well designed hierarchical 
initialization and coefficient update scheme, a new 
model based rate allocation scheme composed of amortization, 
smooth window and consistency control is proposed for a better rate allocation.
Experimental results implementing the proposed
algorithm in the HEVC reference software HM-16.9 show that 
the proposed rate control algorithm is able to achieve an 
average of BDBR saving of 6.09\%, 3.15\% and 4.03\% for random 
access (RA), low delay P (LDP) and low delay B (LDB) 
configurations respectively as compared with the R-$\lambda$ 
model based RC algorithm \cite{li2018lambda} implemented in HM.
The proposed algorithm also outperforms the
state-of-the-art algorithms, while rate control accuracy and 
encoding speed are hardly impacted.
\end{abstract}


\section{Introduction}
\label{sec:intro}
HEVC \cite{sullivan2012overview} is the latest video 
compression standard from ITU and MPEG as the successor to
H.264/AVC \cite{richardson2003h}. It has been widely observed 
that HEVC can save 50\% of the bits on average as 
compared to H.264/AVC while achieving the same visual 
quality, at a cost of much higher encoding complexity \cite{ohm2012comparison}.
Even though video can be encoded using the constant quantization parameter (CQP) mode, 
also known as the non-RC mode, in practical applications, 
rate control is more commonly used to encode an input
video to a target bit rate for bandwidth constrained 
applications while achieving good video 
quality after compression.

Rate control can be generally categorized into two types, 
constant bit rate (CBR) control and average bit rate (ABR) control.
ABR sets a target average bit rate for the entire 
video or every single video segment while allowing the bit rate to 
vary among different parts of the video according to the complexity 
of those parts. On the other side, CBR requires a strictly 
uniform output bit rate for every time period, typically one 
second. At a given bit rate, ABR usually provides a higher quality 
after compression than CBR, as CBR sacrifices a lot in coding 
efficiency for constant bit rates over time. ABR is 
usually used in coding efficiency oriented applications like video 
on demand and video storage, while CBR is often used in
jitter-sensitive applications like video call and satellite based 
video communication. This paper focuses on proposing a novel ABR 
algorithm to improve the coding efficiency of HEVC, so only ABR is 
discussed and tested in this paper.

In general, rate control algorithms need to achieve a high 
bit rate accuracy as measured by bit rate error, while 
achieving good video coding efficiency, which is generally measured 
by BDBR \cite{bjontegaard2008improvements}.
On a high level, rate control consists of two steps, 
\begin{enumerate}
\item allocating the target bit rate across and inside 
frames of the input sequence,
\item selecting proper coding parameters to meet the target 
bit rate with good video quality.
\end{enumerate}

Existing rate control algorithms for HEVC typically use one 
of three rate estimation models, namely the R-Q model \cite{ma2005rate, choi2012rate}, 
the R-$\rho$ model \cite{liu2010low,wang2013rate}, and the R-$\lambda$ model \cite{li2014lambda,li2018lambda}.
These models were designed to predict the output bit rate R after compression 
using features such as the quantization $Q$ in the R-Q model, 
the percentage of zero-valued transformed coefficients $\rho$ 
in the R-$\rho$ model and the Lagrangian multiplier $\lambda$ in the R-$\lambda$ model.

Experiments show that R-$\lambda$ model based rate control algorithms
\cite{li2014lambda, li2018lambda} significantly outperform the R-Q 
model and R-$\rho$ model based algorithms.
The first R-$\lambda$ model based rate control algorithm 
\cite{li2014lambda} is 15\% better in coding efficiency 
than the previous state-of-the-art R-Q model based rate control 
algorithm \cite{choi2012rate} with a nearly halved bit rate error.
As a result, R-$\lambda$ model based algorithms \cite{li2014lambda, 
li2018lambda} have been adopted and integrated in the HEVC reference software.
However, as mentioned in \cite{li2014lambda,li2018lambda},
the coding efficiency of the current R-$\lambda$ model based 
rate control algorithm is still much lower than the CQP mode.
In addition, Wen et al \cite{wen2015r} pointed out 
that the current R-$\lambda$ model based rate control 
algorithm might fail when meeting scene changes.

To improve the performance of R-$\lambda$ model based rate 
control, many algorithms \cite{li2017optimal,xie2015temporal,
wang2016low,he2017efficient,guo2018optimal,wen2015r,
lee2014frame} have been proposed to improve 
rate allocation and/or model coefficients update mechanisms. 
These algorithms achieve a higher coding efficiency than the 
original R-$\lambda$ model based RC algorithm proposed in \cite{li2014lambda,li2018lambda}.
However, those algorithms mainly focus on improving the 
rate allocation and model coefficients update based on the
R-$\lambda$ model without further improving the R-$\lambda$ model.

In this paper, we propose a novel generalized
rate-distortion-$\lambda$ model to better model the 
relationship between rate, distortion and $\lambda$.
The proposed algorithm improves the accuracy of model fitting 
by 56\% over R-$\lambda$ model.
Besides the new model, the well designed hierarchical initialization 
and the model coefficients update scheme, a new model based rate 
allocation scheme is proposed for a better rate allocation.
The rate allocation module includes amortization for I frame, 
smooth window based compensation and consistency control on QP value selection.
Experimental results implementing the proposed 
algorithm in the HEVC reference software HM-16.9 show that 
the proposed rate control algorithm is able to produce
average BDBR savings of 6.09\%, 3.15\% and 4.03\% for the
Random Access (RA), Low Delay P (LDP) and Low Delay B (LDB) configurations respectively as 
compared with the R-$\lambda$ model based RC algorithm in
\cite{li2018lambda}, i.e. the default RC algorithm in HM.
The proposed algorithm also outperforms the state-of-the-art 
rate control algorithms, while
the rate control accuracy and encoding speed are hardly impacted.

The remainder of this paper is organized as follows. Sec. 
\ref{sec:rw} reviews the research on HEVC rate control. Sec. 
\ref{sec:prop} describes the proposed algorithm in detail. Sec. 
\ref{sec:exp} presents experimental results. Section \ref{sec:conclusion} concludes the paper.

\section{Related Work}
\label{sec:rw}

\subsection{Rate Control Models}
\label{sec:model}
In HEVC encoding, the quantization parameter (QP) and the Lagrangian 
multiplier $\lambda$ for rate-distortion optimization (RDO) are two important 
parameters that directly determine output video quality and bit 
rate after compression. QP decides the 
quantization step that is used to quantize the residual
after transform, which determines the distortion of each predicting 
mode as well as the residual after quantization.
$\lambda$, as a function of QP, defines the following RDO 
target function in \cite{sullivan1998rate},
\begin{eqnarray}
J = min(D + \lambda \cdot R),
\end{eqnarray}
where $J$ is also known as the RD cost. It has been widely agreed 
that the following logarithmic relationship \cite{li2012qp} between 
QP and $\lambda$ can achieve the best encoding efficiency statistically.
\begin{equation}
\label{eqn:qp}
QP = c_1 \times ln(\lambda) + c_2,
\end{equation}
where $c_1$ and $c_2$ are variables related to the video 
characteristics and compression performance.
Based on this relationship, rate control algorithms only need 
to decide either QP or $\lambda$. Then various models are used to 
estimate the coding parameters according to the target bit rate.

The R-Q model assumes that the quantization step $Q$ 
has a direct correspondence to the coding complexity and can 
accurately estimate the number of bits consumed using the following
quadratic model \cite{ma2005rate}, 
\begin{eqnarray}
\label{eqn:rq}
R &=& aQ^{-1} + bQ^{-2},\\
QP &=& 12.0 + 6.0 \times log(Q / 0.85),
\end{eqnarray}
where $a$ and $b$ are two parameters related to video 
content that are updated as encoding proceeds.
Choi et al \cite{choi2012rate} proposed a pixel based 
unified R-Q model based rate control algorithm for HEVC, 
which was adopted and implemented in HEVC reference software 
HM-8.0. As discussed, $Q$ can only determine the distortion 
and residual of each predicting mode. However, $\lambda$ decides 
which prediction mode to use, while the output bit rate of the 
current prediction is determined by CABAC. Therefore, the indirect 
relationship between R and Q cannot achieve a high accuracy in rate estimation.
In addition, the non-monotonic quadratic model in R-Q model 
is hard to be accurately updated during the encoding process, 
which would also reduce the coding efficiency.

Another class of rate control algorithms, namely $\rho$ domain
rate control \cite{he2001low,liu2010low,wang2013rate}, assumes a 
linear relationship, shown in Equation (\ref{eqn:rho}), between 
the output bit rate and the percentage of zeros among the 
quantized transform coefficients, denoted as $\rho$.
\begin{eqnarray}
\label{eqn:rho}
R = \theta \cdot (1 - \rho).
\end{eqnarray}
$\rho$ domain rate control algorithms were popular for H.264.
However, HEVC standard introduces a flexible quad-tree coding 
unit (CU) partition scheme and the skip prediction method, leading 
to significant difference in the distribution of zeros after transform and quantization.
In addition, the extra bits consumed by newly introduced
syntaxes also make the linear relationship between $\rho$ and 
output bit rate less accurate as compared with H.264. As a result,
$\rho$ domain rate control is rarely used in HEVC.

As the state of the art, Li et al \cite{li2014lambda} 
proposed a novel R-$\lambda$ model for HEVC, where the 
relationship between distortion and rate is modelled as a 
hyperbolic function in Equation (\ref{eqn:rd}). Accordingly,
the relationship between bit per pixel (bpp) and 
$\lambda$ is also hyperbolic after unit conversion, as given in Equation (\ref{eqn:rlambda}).
\begin{eqnarray}
\label{eqn:rd}
D &=& C R^{-K},\\
\label{eqn:rlambda}
\lambda &=& \alpha bpp^{\beta},
\end{eqnarray}
where $C$, $K$, $\alpha$ and $\beta$ are content related 
parameters. Only $\alpha$ and $\beta$ are needed to be estimated and updated during encoding. 
Similar to the hierarchical structure used in HEVC, $\lambda$
domain rate control algorithms also propose its own hierarchy, 
where frames in the same frame-reference hierarchy share a same set of model coefficients.
During encoding, $\alpha$ and $\beta$ are updated 
using a least mean square based gradient descent scheme.
Experimental results \cite{li2014lambda} show that R-$\lambda$ model 
is able to properly model the relationship 
between distortion and rate with a coefficient of determination ($r^2$) value around 0.995.
Compared with R-Q model based rate control algorithm
\cite{choi2012rate}, the former state of the art, R-$\lambda$ model based rate control 
algorithm \cite{li2014lambda} improves the coding efficiency
by 15.9\% for LD and 24.6\% for RA respectively.
As a result, the R-$\lambda$ model based rate control algorithm 
in \cite{li2014lambda} was adopted and implemented in the HEVC 
reference software HM-10.0. However, experimental results in 
\cite{li2014lambda} also show that the coding efficiency of the
R-$\lambda$ model based algorithm is still far inferior to CQP.

\subsection{Bit Allocation}
\label{sec:ba}
The R-$\lambda$ model in Equation (\ref{eqn:rlambda}) is  
highly effective for output rate estimation and determining the QP 
value to hit a target bit rate. However, the question of 
optimally allocating the total bit rate budget to frames and/or
CUs for a higher video coding efficiency remains open.

Li et al \cite{li2018lambda} proposed a largest coding unit
(LCU)-level separate model based block 
level bit allocation scheme, which achieves a higher rate control 
accuracy and also a higher coding efficiency (2.8\% and 3.9\% 
for LD and RA) than \cite{li2014lambda}. As a result, the algorithm  
was incorporated into HEVC reference software HM-11.0.

As LD configuration and RA configuration are very different from 
each other in rate distribution, some rate allocation algorithms 
were designed to work with only one of the two configurations.
For example, \cite{he2017efficient} was predominantly designed to 
work very well in the RA mode, whereas \cite{li2017optimal} is among 
the best for LD. RA uses a more complex reference hierarchy that 
needs to adapt to input content, so it is generally more
difficult for RC algorithms to work well with RA.

Xie et al \cite{xie2015temporal} proposed a temporal dependent bit 
allocation scheme which allocates bit rate according to
the complexity of each coding tree unit (CTU). Results show a 
coding efficiency improvement of 1.78\% over \cite{li2014lambda} for LD.
Wang \cite{wang2016low} et al proposed a new relationship between 
the distortion and $\lambda$ for a better rate regulation and a 
higher consistency in quality with an average gain of 0.37 dB in PSNR.
Li et al \cite{li2017optimal,guo2018optimal} proposed a 
recursive Taylor expansion method to iteratively 
estimate a close form of the optimal rate allocation, which
improved the coding efficiency by 2.2\% and 2.4\% on average over 
\cite{li2018lambda} for LDP and LDB respectively.

As to RA configuration,
Song et al \cite{song2017new} proposed a group of pictures (GOP) 
level rate allocation scheme to accurately match the HEVC GOP 
coding structure in RA, which achieves a coding efficiency that is 0.2\% higher than \cite{li2018lambda}.
Gong et al \cite{gong2017temporal} proposed a
temporal-layer-motivated lambda domain picture level rate control algorithm to 
better estimate the influence of each layer, which leads to an average 
gain of 3.87\% in coding efficiency as compared with \cite{li2018lambda}.
He et al \cite{he2017efficient} proposed an inter-frame dependency 
based dynamic programming method for frame level bit allocation, 
which improves the coding efficiency by 5.19\% on average for RA 
than \cite{li2018lambda} with an increase of 0.41\% in encoding time .

\subsection{Other methods}
\label{sec:other}
Besides rate allocation, various schemes have been proposed for a 
better RC performance, such as adaptive quantization, new RC models and 
multi-pass encoding.

Adaptive quantization is another mean of bit allocation, which 
usually first uses traditional RC algorithms to decide a central QP 
value. The QP values for frames and CUs are adjusted later.
Tang et al \cite{tang2018hadamard} proposed a Hadamard energy 
feature for adaptive quantization and achieved 3.3\% gain in coding 
efficiency as compared with \cite{li2018lambda}.

Lee et al \cite{lee2014frame} proposed a Laplacian 
probability density function to derive a new model between rate and 
distortion. The coding performance is slightly better than 
\cite{li2014lambda} but worse than \cite{li2018lambda}, so this 
model was not adopted by HEVC reference software.

Besides only using features inside a frame, there are also some 
multi-pass methods, which increases the coding efficiency at a 
cost of higher latency and higher computing complexity.
Wen et al \cite{wen2015r} proposed a pre-compression based double 
update scheme to better handle scene changes, which 
achieves an average gain in PSNR of 0.1dB for common single-scene 
test sequences and up to 4.5dB for complicated multi-scene videos.
The macroblock-tree algorithm proposed in \cite{garrett2009novel}  
was designed to adjust the QP value according to the frequency
at which a block is directly and indirectly referenced.
An extra pass of encoding is required in the macroblock-tree 
algorithm, which leads to a higher latency and a higher complexity.
Based on the macroblock-tree algorithm, Yang et al 
\cite{yang2012source} proposed a low-delay source distortion temporal 
propagation model, which improves the coding efficiency of H.264 
reference software JM by 15\%. Fiengo et al \cite{fiengo2016rate} 
proposed a convex optimization based recursive R-D model, which 
achieves a gain of 12\% in coding efficiency as compared with 
\cite{li2014lambda} with a 10x-50x higher encoding time.
Ropert et al \cite{bichon2019optimal,ropert2017rd} proposed a 
sequential two-pass method for a better rate allocation, which
results in an increase of 16\% in coding efficiency at a cost of an 
average increase of 57\% in encoding time as compared with \cite{li2018lambda}.

\section{The Proposed Algorithm}
\label{sec:prop}
In this section, we describe in detail, the proposed 
generalized rate-distortion-$\lambda$ model, and the proposed 
rate control algorithm based on the new model, including the 
hierarchical initialization, least mean square based update 
scheme and the rate allocation module.

\begin{figure*}[b]
\setcounter{subfigure}{0}
\subfigure[PeopleOnStreet 1600p]{\includegraphics[width=0.32\textwidth]{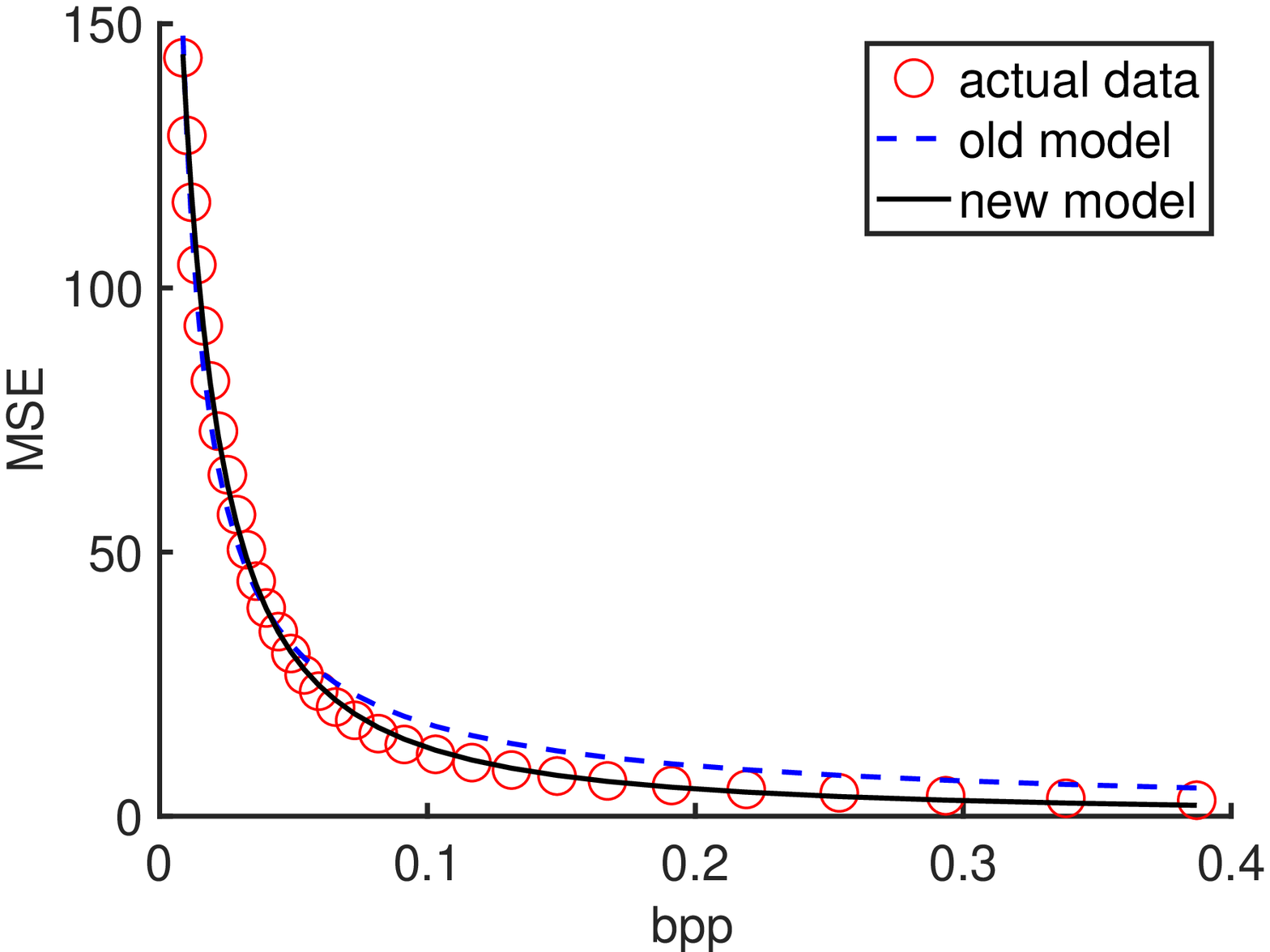}}
\subfigure[PartyScene 480p]{\includegraphics[width=0.32\textwidth]{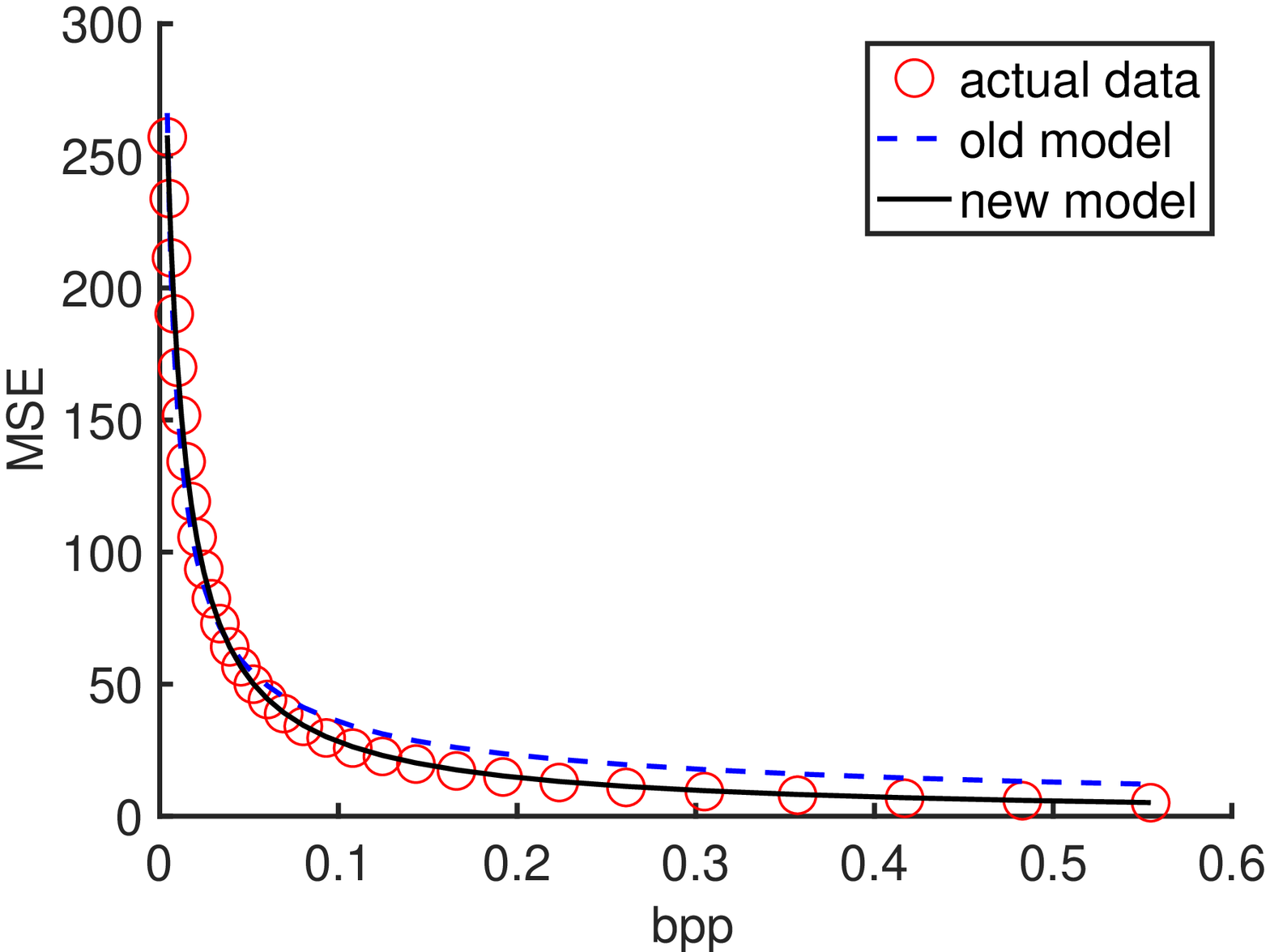}}
\subfigure[BlowingBubbles 240p]{\includegraphics[width=0.32\textwidth]{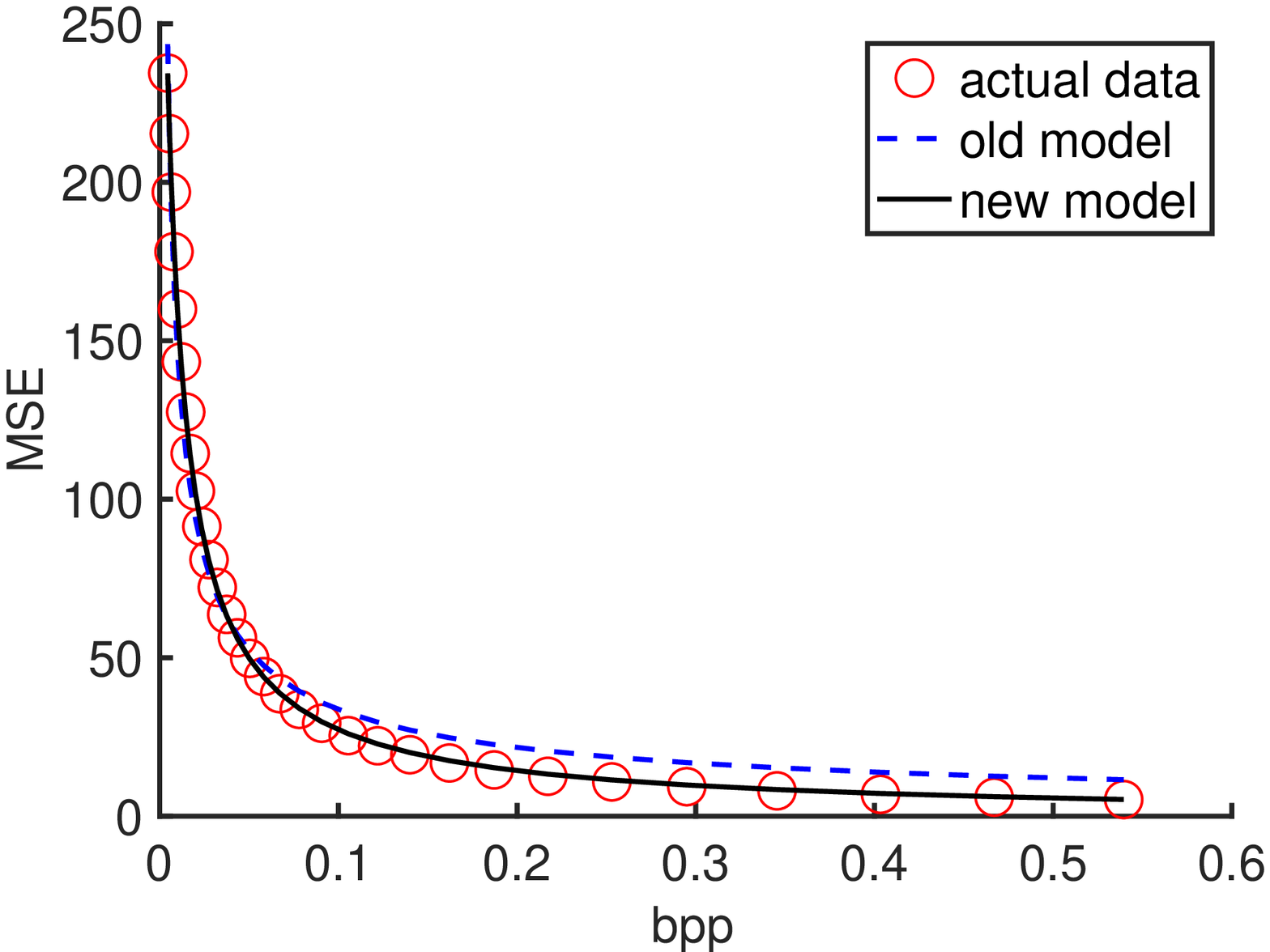}}
\caption{R-D Curves Fit Using the Two Different Models}
\label{fig:fit}
\end{figure*}

\subsection{Rate-Distortion-$\lambda$ Model}
\label{subsec:model}
The core of the R-$\lambda$ model can be found in Equation (\ref{eqn:rd}) and (\ref{eqn:rlambda}).
Though the model performs well in R-D 
fitting, there are two implied assumptions that are invalid under 
border conditions in real world applications, namely, (a) 
infinite bit rate when distortion is zero, (b) 
infinite distortion when bit rate is zero rate.

HEVC provides a lossless encoding mode, where the QP 
value is 4 and the corresponding quantization step is 1.
Therefore, the output rate of lossless encoding is the minimum
bit rate for the video to be encoded without distortion.
Such lossless bit rate is not much greater than output bit rate values from lossy encodings.
For example, the lossless bit rate of Kimono in class B is 3.3 
bpp, and that of FourPeople in class F is 2.2 bpp.
A common target bit rate of lossy encoding usually lies in the 
range between 0.01 bpp and 0.5 bpp, which is only one to two 
orders of magnitude smaller than the rate of lossless encoding.
Therefore, such border cases cannot be ignored in the model.
On the other side, zero rate encoding can be approximated by only 
storing the average value of the video, where the output 
distortion is the variance of the video.

These two boundary cases prove that a more realistic
rate-distortion model must intercept both the rate 
axis and the distortion axis. This is, however, not true for
the R-$\lambda$ model, which manifests as loss 
of model accuracy, especially for very low/high bit rates.

According to the discussion above, the proposed model is given as follows,
\begin{eqnarray}
\label{eqn:newrd}
D &=& max(0, C(R+B)^{-K} - T)\\
\label{eqn:approxnewrd}
&\approx& C(R+B)^{-K} - T,
\end{eqnarray}
where $C$ and $K$ are parameters similar to their name sakes 
in the traditional R-$\lambda$ model that represent the basic 
characters of the video. According to the results in 
\cite{li2014lambda}, $K$ is usually around 1.
$B$ is the parameter describing the interception on the distortion axis. 
The distortion of zero bit rate encoding can be described by
$CB^{-K}$, which equals to the variance of the video.
As a result, $B$ is usually much smaller than a 
typical target bit rate. $T$ is the parameter for the 
interception on the rate axis, which equals to 
$C(R_{lossless} + B)^{-K}$ and usually is one to two orders 
of magnitude smaller than typical output distortion.
The model can be simplified to Expression 
(\ref{eqn:approxnewrd}) if only lossy encoding is considered.

The modelling performance between rate and distortion, i.e. how 
close the model could fit the regression curve to the actual data, 
determines how accurately the model could possibly be updated.
Therefore, a fitting experiment was conducted to evaluate the expressive 
power of the proposed new model in describing the relationship 
between rate and distortion, where test sequences in 
HEVC common test condition \cite{bossen2013common} (class A to E, 20 
videos in total) were encoded using the CQP mode with QP values from 4 to 51 tested to cover 
all possible output rates. When QP is set to 4, the quantization step 
is 1 in encoding, which represents lossless encoding and 
also the highest possible output rate. 51 is the greatest QP 
value that is allowed in HEVC, which leads to a lowest 
possible output rate. For each video sequence, the relationship between distortion 
(measured by mean squared error, MSE) and rate (measured by
bpp) were fit using the rate-distortion model \cite{li2014lambda}
in Equation (\ref{eqn:rd}) and the proposed 
model defined in Equation (\ref{eqn:approxnewrd}).

\begin{table}[t]
    \centering
    \scriptsize
    \caption{RD Curves Fitting Performance}
    \label{tab:fit}
    \begin{tabular}{cccccccc}
    \toprule
     \multirow{2}[2]{*}{QP Range} & \multirow{2}[2]{*}{Model} & \multicolumn{3}{c}{$r^2$} & \multicolumn{3}{c}{RMSE} \\
     \cmidrule{3-8}
       &   & worst & avg & std & worst & avg & std \\
    \midrule
    \multirow{2}[2]{*}{4...51} & \cite{li2014lambda}     & 0.9504 & 0.9925 & 0.0106 & 11.24 & 3.03 & 2.91 \\
    & Proposed     & 0.9777 & 0.9966 & 0.0055 & 7.62 & 1.32 & 1.62 \\
    \midrule
    \multirow{2}[2]{*}{4...22} & \cite{li2014lambda}     & 0.9046    & 0.9831  &  0.0242 &   1.59  &  0.32  &  0.36 \\
    & Proposed     & 0.9763  &  0.9940  &  0.0074   & 0.36   & 0.15 &   0.11\\
    \midrule
    \multirow{2}[2]{*}{17...37} & \cite{li2014lambda}     & 0.8904   & 0.9880 &   0.0252  &  9.89  &  1.47  &  2.10 \\
    & Proposed     & 0.9603 &   0.9941  &  0.0108 &   2.65   & 0.65  &  0.72\\
    \midrule
    \multirow{2}[2]{*}{32...51} & \cite{li2014lambda}     & 0.9818    & 0.9978  &  0.0040  &  8.24  &  2.14  &  2.37 \\
    & Proposed     & 0.9880  &  0.9992  &  0.0027  &  6.90   & 0.76  &  1.47 \\
    \bottomrule
    \end{tabular}%
\end{table}%

Table \ref{tab:fit} gives the fitting results  
of the model used in R-$\lambda$ model and the proposed model 
with coefficient of determination ($r^2$) and root mean square 
error (RMSE) used as the metrics. $r^2$ describes how close 
the actual data is to the fitted regression curve. RMSE is the 
standard deviation of the prediction residues. A higher $r^2$ 
value or a lower RMSE value for a same set of data suggests a better fitting.

In the experiment, four QP ranges (namely, full QP range: 4...51, low 
QP range: 4...22, middle QP range: 17...37, high QP range: 32...51) were tested to
compare the expressive powers of two models for different QP 
ranges. Each of the sub-ranges contains around 20 points to guarantee 
a similar difficulty in fitting. As shown in Table \ref{tab:fit}, 
for the full QP range (4...51), the proposed model is able to 
improve $r^2$ value from 0.9925 to 0.9966 on average, with RMSE 
reduced from 3.03 to 1.32, i.e. a 56\% reduction.
It should be noted that the $r^2$ values in the proposed experiment 
are lower than the results in \cite{li2014lambda} because 48
data points (QP from 4 to 51) were used in our experiment, while 
only four points (QP in \{22, 27, 32, 37\}) were used in \cite{li2014lambda}.
Furthermore, the proposed model is intended for both the average 
and worst cases for full QP range, while the much lower standard deviations 
(std) of $r^2$ and RMSE show that the proposed model is more 
robust than the model in \cite{li2014lambda} when handling different kinds of videos.

In addition, the results in Table \ref{tab:fit} prove that the 
proposed model outperforms the R-$\lambda$ model for all 
three sub-ranges of QP values. The R-$\lambda$ 
model sometimes produces very low $r^2$ values for 
low QP values. As discussed above, a common target 
bit rate of lossy encoding is only one to two orders of magnitude 
smaller than the rate of lossless encoding. As a result, the 
proposed model greatly benefits from the refinement on the 
zero-distortion boundary case and therefore produces a much higher 
and stabler fitting performance. 
On the other side, the distortion of zero rate encoding, i.e. the 
variance of the video, is usually much greater than the distortion 
after compression, so the gain in $r^2$ for the high QP range due 
to the zero-rate boundary case is expected to be smaller than that 
for the low and middle QP ranges. The R-$\lambda$ model is already 
very accurate for the high QP range, while the proposed model still 
provides a similar improvement in RMSE for the high QP range as compared with other ranges.
It should be noted that $r^2$ and RMSE both increase for the high 
QP range as compared with the low and middle QP ranges because 
RMSE is proportional to distortion. So RMSE increases for the high 
QP range due to a much greater distortion in spite of a higher $r^2$ value.

Fig. \ref{fig:fit} gives three examples of the RD curves 
fitted using the two models. The original data points are plotted 
in distinct red circles, while the curves fitted using the old model 
and the proposed model are plotted using blue dashed line and black
solid line respectively. It needs to be noted that the 
model coefficients were estimated using all data points in the full QP range, while 
only a part of data points (QP 15 to 44) is plotted in Fig. 
\ref{fig:fit} to prevent the large range making the plot hard to discern.  
As can be seen from the figure, the R-$\lambda$ model performs well 
for high QP values and starts to deteriorate with a lower QP value, while the proposed model is 
able to accurately predict the RD relationship for all cases.

Based on the proposed rate-distortion model defined in Equation 
(\ref{eqn:approxnewrd}), the proposed R-D-$\lambda$ model can be 
derived as follows,
\begin{eqnarray}
\label{eqn:lambdadef}
\lambda&=&-\frac{\partial D}{\partial R}=CK(R+B)^{-K-1}\\
\label{eqn:lambdabpp}
&=&\alpha(bpp+\gamma)^\beta,\\
\label{eqn:bpplambda}
bpp&=&(\frac{\lambda}{\alpha})^{1/\beta}-\gamma,\\
\label{eqn:model_alpha}
\alpha &=& CK,\\
\label{eqn:model_beta}
\beta &=& -K-1,\\
\label{eqn:model_gamma}
\gamma &=& \frac{B}{W\times H \times Fr},
\end{eqnarray}
where $\lambda$ is the opposite number of the derivative between 
distortion and rate. Equation (\ref{eqn:lambdabpp}) and 
(\ref{eqn:bpplambda}) provide another two interpretations of 
Equation (\ref{eqn:lambdadef}) with the unit of rate converted from bits to bpp.
$\alpha$, $\beta$ and $\gamma$ are the short codes of the 
video characters related parameters with definitions given in Equation 
(\ref{eqn:model_alpha})-(\ref{eqn:model_gamma}). These three 
parameters are updated using the algorithm proposed in Sec. 
\ref{subsec:update} as coding proceeds. $W$, $H$ and $Fr$ are the 
width, height and frame rate of the video.

Based on the model definition introduced above, $\lambda$ can be 
calculated according to the target bit rate.
And the value of QP can be calculated using the logarithmic function 
defined in Equation (\ref{eqn:qp}).
The values of $c_1$ and $c_2$ in Equation (\ref{eqn:qp}) were 
determined through a tuning experiment where various values were 
tested to cooperate with the remaining parts of the proposed algorithm.
It was found that the following QP-$\lambda$ relationship 
provided the best coding efficiency for the HEVC common test 
condition, and therefore is used in the proposed algorithm.
\begin{eqnarray}
QP = round(4.3\times ln(\lambda) + 14.6).
\label{eqn:newqp}
\end{eqnarray}

\subsection{Hierarchical Initialization}
\label{subsec:hier}

Similar to \cite{li2014lambda}, in the proposed algorithm, 
frames of a same reference hierarchy share a same set of model coefficients, 
which is updated after every encoding of frame.
All CUs inside a frame share the model of that frame 
without separate maintenance.
The update mechanism of the model coefficients is introduced in 
Sec. \ref{subsec:update}.

In \cite{li2014lambda}, the initial values of $\alpha$ and
$\beta$ are set to 3.2003 and -1.367 respectively, which are 
the averaged fitted values using the sequences in the HEVC 
common test condition. It is agreed that the initial values of 
model coefficients do not have an overall significant influence on 
the coding efficiency as long as model coefficients can be properly updated during encoding.
However, it is still beneficial to set the initial values 
according to the reference frame hierarchy.

\begin{table}[t]
\centering
\caption{RA Coding Structure in HEVC}
\label{tab:rastruct}
\begin{tabular}{cccccc}
\toprule
FrameNum & POC & Level & Ref Num & QP Offset & $\lambda$ Multiplier \\ 
\midrule
1 & 8 & 1 & 3 & 1 & 0.442 \\ 
2 & 4 & 2 & 3 & 2 & 0.3536 \\  
3 & 2 & 3 & 4 & 3 & 0.3536 \\ 
4 & 1 & 4 & 4 & 4 & 0.68\\ 
5 & 3 & 4 & 4 & 4 & 0.68 \\ 
6 & 6 & 3 & 3 & 3 & 0.3536 \\ 
7 & 5 & 4 & 4 & 4 & 0.68 \\ 
8 & 7 & 4 & 4 & 4 & 0.68 \\ 
\bottomrule
\end{tabular} 
\vspace{.3cm}
\caption{LD Coding Structure in HEVC}
\label{tab:ldstruct}
\begin{tabular}{cccccc}
\toprule
FrameNum & POC & Level & Ref Num & QP Offset & $\lambda$ Multiplier \\ 
\midrule
1 & 1 & 3 & 4 & 3 & 0.4624 \\ 
2 & 2 & 2 & 4 & 2 & 0.4624 \\  
3 & 3 & 3 & 4 & 3 & 0.4624 \\ 
4 & 4 & 1 & 4 & 1 & 0.578\\ 
\bottomrule
\end{tabular} 
\end{table}

\begin{figure}[b]
\center
\includegraphics[width=0.50\textwidth]{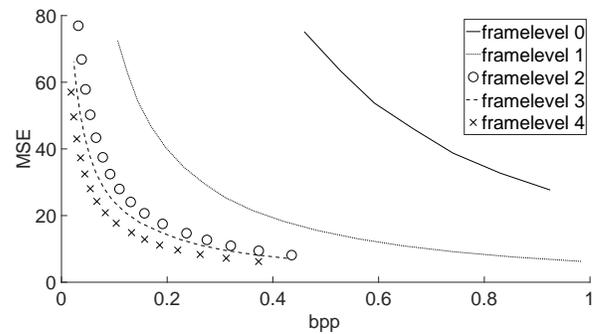}
\caption{R-D Curves of Different Frame Levels}
\label{fig:framelevel}
\end{figure}

A hierarchically structured GOP mechanism with frames at different 
levels of the hierarchy using different numbers of reference 
frames and different QP values was proposed in HEVC.
In general, a smaller QP value is used for frames of a 
lower level, which are referenced more and reference other frames less.
The hierarchical structures of HEVC (RA in Table 
\ref{tab:rastruct}, LD in Table \ref{tab:ldstruct}) were 
carefully designed and fine-tuned for a statistically optimal 
coding efficiency under the HEVC common test condition.
Intra-picture coded frames (I frames) belong to frame level 0, 
while inter-picture coded frames (P frames and B frames) are 
categorized into frame level 1 to 4 in RA.
A lower frame level represents a higher importance in the 
reference structure and a lower dependence on other frames in 
coding and therefore a usually lower coding efficiency.
Fig. \ref{fig:framelevel} gives the RD curves of different 
frame levels of ``PartyScene'' from class C, which was encoded by 
HM-16.9 using RA configuration with QP values from 22 to 37.
As shown in Fig. \ref{fig:framelevel}, for a given 
output quality, frames of frame level 4 only require about 25\% of 
the bit rate that is needed by frames of frame level 1. Therefore, 
it is not proper to set the initial values of all frame levels to the same value.

To derive a close form of the optimal relationship between model parameters
and frame level, it is assumed that when a video sequence is encoded to a bit rate of 
$R$ by encoder A and a certain quality, encoder B would encode that video into a similar 
quality if the output bit rate is $dR$, where $d$ is the relative 
coding efficiency between encoder B and encoder A over a range of bit rates.
This assumption is widely used in many encoder evaluation 
schemes, such as the BDBR metric \cite{bjontegaard2008improvements}. 
In this paper, it is assumed that the linear scaling of 
difference in coding efficiencies between two encoders is applicable to
the coding efficiencies of two frame levels as well. 
The distortions when encoding a frame at two different 
frame levels ($i$ and $j$) by the same encoder can be modelled by
\begin{eqnarray}
D &=& C_i (R + B_i)^{-K_i} - T_i\\
&=& C_j (dR + B_j)^{-K_j} - T_j,
\end{eqnarray}
where $d$ is the relative coding efficiency between level $j$
and $i$. $T_i$ is the feature same in Expression
(\ref{eqn:approxnewrd}), which equals to $C_i(R_{i, lossless} 
+ B_i)^{-K_i}$ and is one to two orders smaller than that in lossy 
encoding cases. The difference between $T_i$ and $T_j$ 
is much smaller than common distortion and therefore can be ignored.
As introduced earlier, $B$ is usually much smaller than practical bit rates so that Expression 
(\ref{eqn:multilevel}) and (\ref{eqn:rep}) can be approximated 
using Taylor expansion. According to the results in 
\cite{li2014lambda}, $K$ is usually around 1 so that $K$ will 
not affect the order of magnitude of $\frac{B}{R}$.
So $\frac{B_i B_j}{R^2}$ in Expression 
(\ref{eqn:ag}) is much smaller than 1, and 
therefore can be ignored to get the final approximation.
The entire approximation is described as follows,
\begin{eqnarray}
\label{eqn:multilevel}
C_i (R + B_i)^{-K_i} &\approx& C_j (dR + B_j)^{-K_j}\\
C_i R^{-K_i}(1-\frac{K_i B_i}{R}) &\approx& C_j (dR)^{-K_j}(1-\frac{K_j B_j}{dR})\\
\label{eqn:rep}
R^{{K}_{j}-{K}_{i}} &\approx & \frac{C_{j}}{C_i}{d}^{-{K}_{j}}\frac{1-\frac{K_j B_j}{dR}}{1-\frac{K_i B_i}{R}}  \\
\label{eqn:ag}
 &\approx &   \frac{C_{j}}{C_i}{d}^{-{K}_{j}}(1-\frac{K_i K_j B_i B_j}{dR^2}) \\
&\approx &  \frac{C_{j}}{C_i}{d}^{-{K}_{j}}
\label{eqn:approx}
\end{eqnarray}
To ensure Expression (\ref{eqn:approx}) roughly holds 
for a wide range of rates to satisfy the assumption, $K_i$ 
and $K_j$ must be very close to each other, while the $C$ values 
for different frame levels roughly follows a reciprocal 
relationship of the relative coding efficiency of each frame level.
\begin{eqnarray}
\frac{C_{j}}{C_i} \approx {d}^{K} \approx d
\end{eqnarray}

\begin{table}[bt]
\centering
\caption{$C$ for Different Frame Levels in RA}
\vspace{0.2cm}
\label{tab:C}
\begin{tabular}{ccccc}
\toprule
Frame Level & 1 (P frame) & 2 & 3 & 4\\ 
\midrule
Ref Distance & 8 & 4 & 2 & 1 \\
\midrule
$C$ &  4.180   & 2.905  &  1.892 &   1.020 \\
\bottomrule
\end{tabular} 
\end{table}

Wang et al \cite{wang2007impact} concludes that coding 
efficiency decreases as a logarithmic function of the 
distance between the current frame and the reference, which is 
highly consistent with the fitted results of the RD curves of 
different frame levels given in Table \ref{tab:C}. It should be 
noted that the logarithmic relationship is not a good fit for frame level 1 (P frames), as 
frames in level 1 only allows uni-directional 
prediction, while frames of higher frame levels can be encoded 
using bi-directional prediction that improves the coding efficiency.

Moreover, the hierarchical initialization scheme for $\gamma$
can be obtained using the zero rate encoding case, where the 
theoretical maximum distortion is the variance, which is 
similar for frames within one scene.
Therefore, the hierarchical initialization scheme for $\gamma$ 
can be derived as follows,
\begin{eqnarray}
C_i {B_i}^{-K} &\approx& C_j {B_j}^{-K} \\
(\frac{B_j}{B_i})^K &\approx& \frac{C_j}{C_i} \quad \approx d^K \\
\frac{\gamma_j}{\gamma_i} &\approx& \frac{C_j}{C_i} \quad \approx d
\end{eqnarray}

In summary, the model coefficients of different frame levels
are initialized hierarchically, while frames of the same frame 
level share the same set of model coefficients.
For the RA configuration, all frame levels share the 
same $\beta$ value of -1.35, while $\alpha$ and $\gamma$ 
follow a fixed proportional relation of [4.2:3:2:1] from the results in Table \ref{tab:C} for 
frame level 1 to 4 with center values (frame level 2) set to 4.4
and 0.005 for $\alpha$ and $\gamma$ respectively.
For the LD configuration, $\beta$ is also set to -1.35,
while $\alpha$ and $\gamma$ are set to 2.4 and 0.005 for frame level 1 to 3.
0.005 is the averaged fitted values of $\gamma$ using the 
sequences in HEVC common test condition.
A $\gamma$ value of 0.005 (i.e. 140kbps for 720p@30fps and 60kbps for 
480p@30fps) is usually much lower than practical target bit rates.
It is still possible that $\gamma$ is comparable with the 
target bit rate for very simple videos. In those cases, 
the relationship that $B$ is much 
smaller than target bit rate will not sustain and
the approximations like Expression (\ref{eqn:rep}) will fail. 
To ensure this relationship, the initial $\gamma$ is 
clipped by 0.1 times of the target bit rate as the upper bound.

\subsection{Model Parameters Update Scheme}
\label{subsec:update}
In the proposed rate control algorithm, frames of a same 
frame level share a same set of model coefficients, which is 
updated after every encoding of a frame. Unlike the LCU-level 
separate model proposed in \cite{li2018lambda}, the proposed 
algorithm only allows separate models for different frame 
levels. CUs inside a frame share the model of this frame, 
which will not be updated until this frame is completely encoded.
In the proposed algorithm, each frame 
is first allocated with a target bit rate, denoted as 
$bpp_{0}$. The estimated coding parameter $\lambda$
(denoted as $\lambda_0$) can be calculated using
\begin{equation}
\label{eqn:est}
\lambda_{0}=\alpha_{i}(bpp_{0}+\gamma_{i})^{{\beta}_{i}},
\end{equation}
where $i$ is the frame level of the current frame. The 
corresponding QP value ($QP_0$) can be calculated using 
Equation (\ref{eqn:newqp}).

After the current frame is encoded, the output bit 
rate ($bpp_1$) can be observed, which is used to 
update the proposed R-D-$\lambda$ model using 
the least mean square (LMS) update rule.
The $\lambda$ estimation for hitting a target bit rate can be 
considered a regression problem, where $bpp_0$ is the input 
variable, and $\lambda_0$ is the estimated output.
Then ($bpp_1$, $\lambda_0$) becomes an actual data point of 
the model, while $\lambda_1$ is the estimated biased output, 
which can be calculated as follows.
\begin{equation}
\lambda_{1}=\alpha_{i}(bpp_{1}+\gamma_{i})^{{\beta}_{i}}.
\end{equation}
To make the power function easy to be updated in gradient 
descent, a squared logarithmic error ($e^2$) is used as follows,
\begin{eqnarray}
e^2 &=& \frac{1}{2}(ln\lambda_{0}-ln\lambda_{1})^2,\\
ln\lambda_{1} &=& ln\alpha_{i} + \beta_{i}ln(bpp_{1}+\gamma_{i})
\end{eqnarray}
The derivatives between $e^2$ and $\alpha$, $\beta$ as well as
$\gamma$ can be calculated as follows,
\begin{eqnarray}
\frac{\partial{e^2}}{\partial{\alpha_i}} &=& \frac{\partial{e^2}}{\partial{ln\lambda_1}}\frac{\partial{ln\lambda_1}}{\partial{ln\alpha_i}}\frac{\partial{ln\alpha_i}}{\partial{\alpha_i}}\nonumber\\
 &=& -(ln{\lambda}_{0} - ln{\lambda}_{1})\frac{\partial{ln{\lambda}_{1}}}{\partial{ln\alpha_i}}\frac{\partial{ln\alpha_i}}{\partial{\alpha_i}}\nonumber \\
&=&-(ln{\lambda}_{0} - ln{\lambda}_{1})\frac{1}{\alpha_i}\\
\frac{\partial{e^2}}{\partial{\beta_i}} &=& -(ln{\lambda}_{0} - ln{\lambda}_{1})\frac{\partial{ln{\lambda}_{1}}}{\partial{\beta_i}} \nonumber\\
&=&-(ln{\lambda}_{0} - ln{\lambda}_{1}) ln(bpp_{1} + \gamma_{i}) \\
\frac{\partial{e^2}}{\partial{\gamma_i}} &=& -(ln{\lambda}_{0} - ln{\lambda}_{1})\frac{\partial{ln{\lambda}_{1}}}{\partial{\gamma_i}} \nonumber \\
&=& -(ln{\lambda}_{0} - ln{\lambda}_{1})\frac{\beta_i}{bpp_1 + \gamma_i},
\end{eqnarray}

Based on the LMS update rule, the model coefficients can be 
updated using the updating strength set ($\sigma_{\alpha}$,
$\sigma_{\beta}$, $\sigma_{\gamma}$) as follows. The symbols 
with a dot symbol above are the coefficients after update.
\begin{eqnarray}
\label{eqn:thetaupdate}
\dot{\alpha_i}&:=&\alpha_i + \sigma_{\alpha}(ln{\lambda_0}-ln{\lambda_1})\frac{1}{\alpha_i},\\
\dot{\beta_i}&:=&\beta_i+\sigma_{\beta}(ln{\lambda_0}-ln{\lambda_1})ln(bpp_1 + \gamma_i),\\
\dot{\gamma_i}&:=&\gamma_i+\sigma_{\gamma}(ln{\lambda_0}-ln{\lambda_1})\frac{\beta_i}{bpp_1 + \gamma_i}.
\end{eqnarray}

In the proposed algorithm, the initial update strengths of
$\alpha$, $\beta$ and $\gamma$ are set to 0.05, 0.2 and 
0.000001 times of the target bpp, which gradually 
decrease during the encoding process using a decay scheme.
Inspired by the decreasing learning rate schemes that are widely 
used in deep learning applications, an exponential decay scheme 
is used in the proposed algorithm for a better convergence 
and noise reduction. In the proposed algorithm, each frame 
level maintains a set of model coefficients as well as a 
decay value, which is multiplied by 0.99 after a 
frame of that level is encoded.
Furthermore, the efficient scene change detection algorithm 
proposed in our previous work \cite{tang2018hadamard} is 
included in the proposed algorithm. If a scene change is 
observed, the model parameters are reset to initial 
values and the decay value is reset to 1. It should be noted that 
the videos in the HEVC common test condition do not include scene 
changes, so the scene change detection module was 
disabled in the experiment.

\subsection{Rate Allocation}
\label{subsec:ratealloc}
The rate allocation scheme in the proposed algorithm can be separated 
into two levels, GOP level and picture level. CUs 
inside a frame share a same set of coding parameters for
a spatially consistent output quality. The rates mentioned in this section are all measured in bpp.  

As a special case, I frames are usually very different from 
inter-picture coded frames in terms of R-D  relationship, 
so there is usually a special rate control module 
for I frames. Similar as \cite{li2014lambda,li2018lambda}, the 
proposed rate allocation module first treats I frames as
inter-picture coded frames and obtain a target bit rate for that 
frame. Then the algorithm in \cite{karczewicz2013jctvc} is 
used to refine the target bit rate and get a new target bit rate 
as well as a new set of coding parameters.
As I frames usually consume a bit rate that is much higher than the 
average target bit rate, Li et al \cite{li2018lambda} proposed a smooth 
window scheme to mitigate the overflow in bit rate consumption, where 
the rate overflow is compensated in the next smooth window 
(typically 40 frames) at any instant moment. The smooth window 
scheme works well for most cases, but it may fail in the cases with 
excessive bit consumption. In addition, the compensation will be 
unbalanced if the planned intra period and the length of smooth 
window are not aligned.

To solve this problem, an amortization and smooth window joint 
scheme with restriction on maximum bit rate for I frames is designed 
in the proposed rate control algorithm.
``Amortization'' is a process where the ``debt'', i.e. the excessive 
consumption of bit rate caused by I frame encoding, is paid off 
(i.e. compensated) by the remaining non-I frames within the 
current intra period. 
For example, an I frame (frame number $i$) is first 
considered as a P frame and allocated 
with a target bit rate ($R_{i0}$). $R_{i0}$ is refined 
using the algorithm in \cite{karczewicz2013jctvc} into a new target 
bit rate $R_{i1}$, which is restricted to be 
not greater than half of the target rate 
consumption of the intra period. After encoding, the I frame is 
encoded into a different number of bits $R_{i2}$. The rate to 
be recorded (denoted as $R_{i3}$, which is $R_{i0}$ for I 
frames and $R_{i2}$ for non-I frames) is used in the smooth 
window module, while the overhead of I frames (i.e. $R_{i2}-R_{i0}$) 
is amortized by the remaining non-I frames within the 
current intra period as follows,
\begin{eqnarray}
R_{am} = \frac{R_{i2}-R_{i0}}{IntraPeriod - 1},
\end{eqnarray}
where $R_{am}$ is the averaged amortized rate compensation for 
each frame, and $IntraPeriod$ is the length of the current intra 
period. It needs to be noted that I-frames overhead is 
amortized in a weighted manner rather than in a uniform way. Details  
are introduced in the frame level rate allocation part.
$R_{am}$ is an averaged compensation to make the calculation 
easier. On top of the amortization scheme, the smooth window 
module proposed in \cite{li2018lambda} only accumulates and 
compensates the overflow caused by non-I frames.

Given the current status of amortization, the 
proposed GOP level bit allocation first deducts the 
target amortization from the average target bit rate.
Then the overflow caused by non-I frames is compensated uniformly for each GOP 
within a smooth window, which is set to 40 frames in the proposed algorithm. According to the
description, the target bit rate of the current GOP (${R}_{GOP}$) 
can be calculated as follows.
\begin{eqnarray}
R_{of} &=& \sum_{i=0}^{N-1}{(R_{i3} - R_{i0})}\\
{R}_{GOP} &=& ({R}_{avg} - R_{am} - \frac{R_{of}}{SW}){N}_{GOP}.
\end{eqnarray}
$R_{of}$ is the accumulated non-I rate overflow of the $N$ 
frames that have been encoded. $SW$ is the length of the 
smooth window. $R_{avg}$ is the average target bit rate, and 
${N}_{GOP}$ is the length of the current GOP.

Within the GOP, the target bit rate for each frame is
allocated in a weighted manner, aiming at a sensible 
hierarchical quality distribution. As the output bit rate can be 
estimated by Equation (\ref{eqn:bpplambda}), the optimal frame 
level rate allocation can be considered as an optimal central
$\lambda$ selection problem, which is introduced as follows,
\begin{eqnarray}
\label{eqn:opt}
\underset{\lambda}{min}(abs(\sum_{i=0}^{N_{GOP}-1}{max(R_{i0}, minRate)} - R_{GOP})),\\
\label{eqn:opt2}
R_{i0} = (\frac{\lambda \omega_i}{\alpha_i})^{1/\beta_i} - \gamma_i,
\end{eqnarray}
where $\alpha_i, \beta_i, \gamma_i$ are the model 
coefficients for frame $i$. $\lambda$ is the central $\lambda$ 
value to be solved, and $\omega_i$ is the $\lambda$ 
multiplier for each frame. $abs(.)$ is the absolute value function.
Same as \cite{li2014lambda}, $minRate$ of one frame is set to 
100 bits as the minimum achievable number of bits in the proposed algorithm.
Similar to the hierarchical scheme introduced in Sec. 
\ref{subsec:hier}, frames of a same frame level share the same 
value of $\omega$.

Based on Equation (\ref{eqn:lambdadef}), the relationship 
between D and $\lambda$ can also be interpreted in the forms of 
Equation (\ref{eqn:dlambda1}) and (\ref{eqn:dlambda2})
\begin{eqnarray}
\label{eqn:dlambda1}
(\frac{D}{C})^{\frac{K+1}{K}} &=& R^{-K-1} = \frac{\lambda}{CK},\\
\label{eqn:dlambda2}
D^{\frac{K+1}{K}} &=& \frac{\lambda}{K}C^{\frac{1}{K}},\\
\label{eqn:dlambda_approx}
D^2 &\approx & \lambda C.
\end{eqnarray}
Expression (\ref{eqn:dlambda_approx}) is approximated from 
Equation (\ref{eqn:dlambda2}) based on the fact that $K$ is 
usually close to 1, which was mentioned in \cite{li2014lambda}.
Table \ref{tab:C} shows that the relationship among the $C$ 
coefficients for different frame levels of RA follows a 
proportional relationship of [4.2:3:2:1]. Therefore, a reciprocal 
relationship of $\lambda$ would roughly produce a similar output 
quality for frames of different frame levels.
As the hierarchical structure was designed to encode a more 
important frame into a better quality, the proposed rate 
allocation algorithm uses a relationship of [1:2.5:4.5:10] for 
the $\omega$ values for different frame levels in RA. After the
$\omega$ values are specified, the optimization in Expression 
(\ref{eqn:opt}) can be solved using an iterative binary search. The 
corresponding $\omega$ values for LD are set to [1:4:5].
The fixed values above were selected through an experiment, which
tried various values to cooperate with the remaining parts of the 
proposed algorithm for a higher coding efficiency.

After the frame level coding parameters are determined, 
CUs inside a frame share the same set of coding parameters, 
which may slightly reduce the accuracy of rate control but 
improve coding efficiency and spatial consistency in output quality.

\subsection{Consistency Control}
\label{subsec:consist}
After $\lambda$ value is specified, QP can be calculated 
using Equation (\ref{eqn:newqp}).
To guarantee the output quality to be consistent over time,
$\lambda$ and QP must not change significantly. 
Therefore, a constraint on the maximum QP difference between 
different frames is used.
The maximum QP difference between two consecutive 
frames of a same frame level is set as 3, while the maximum 
QP difference between two consecutive encoded frames is 10.

\begin{table*}[htbp]
  \footnotesize
  \centering
  \caption{RD Performance for RA Configuration, CQP as Anchor}
  \label{tab:ra}%
    \begin{tabular}{lcccccccccccccc}
    \toprule
    \multirow{2}[4]{*}{Clip} & \multicolumn{2}{c}{\cite{choi2012rate}} & \multicolumn{2}{c}{\cite{li2014lambda}-Frame} & \multicolumn{2}{c}{\cite{li2014lambda}-LCU} & \multicolumn{2}{c}{\cite{li2018lambda}-Frame} & \multicolumn{2}{c}{\cite{li2018lambda}-LCU} & \multicolumn{2}{c}{\cite{he2017efficient}} & \multicolumn{2}{c}{Proposed} \\
\cmidrule{2-15}          & BDBR & $\Delta$R & BDBR & $\Delta$R & BDBR & $\Delta$R & BDBR & $\Delta$R & BDBR & $\Delta$R & BDBR & $\Delta$R & BDBR & $\Delta$R \\
    \midrule
    A\_NebutaFestival & 9.55  & 0.62  & 14.20 & 0.88  & 14.78 & 0.57  & 4.06  & 0.43  & 2.97  & 0.20  & -2.94 & 1.53  & 3.74  & 0.79 \\
    A\_PeopleOnStreet & 44.35 & 0.74  & 41.95 & 0.96  & 50.13 & 0.26  & 18.13 & 0.05  & 24.91 & 0.01  & 22.30 & 1.22  & 1.57  & 0.27 \\
    *A\_SteamLoco & 202.06 & 0.77  & 62.31 & 2.96  & 54.93 & 3.05  & 44.76 & 3.20  & 46.29 & 3.69  & 43.65 & 2.46  & 45.57 & 3.91 \\
    A\_Traffic & 65.00 & 1.49  & 6.03  & 0.60  & 6.06  & 0.52  & 1.20  & 0.77  & 4.57  & 0.87  & 0.55  & 0.56  & 0.77  & 1.51 \\
    B\_BasketballDrive & 33.22 & 0.87  & 4.65  & 0.65  & 7.78  & 0.61  & 2.00  & 0.67  & 6.60  & 0.57  & -0.26 & 0.43  & -1.84 & 0.43 \\
    B\_BQTerrace & 57.54 & 0.40  & 3.74  & 0.88  & 5.76  & 0.67  & 3.88  & 1.44  & 7.44  & 0.93  & 5.01  & 2.11  & -0.37 & 1.79 \\
    B\_Cactus & 76.00 & 0.53  & 0.86  & 0.09  & 1.60  & 0.04  & 1.71  & 0.06  & 3.18  & 0.03  & -3.13 & 1.60  & -2.48 & 0.03 \\
    B\_Kimono & 44.72 & 1.30  & 8.86  & 0.13  & 10.18 & 0.07  & 8.08  & 0.23  & 9.86  & 0.45  & 3.81  & 0.83  & 8.91  & 0.41 \\
    B\_ParkScene & 54.72 & 0.96  & 2.95  & 0.13  & 5.28  & 0.38  & 2.74  & 0.31  & 3.14  & 0.03  & -3.18 & 1.09  & 0.01  & 0.68 \\
    C\_BasketballDrill & 58.56 & 0.50  & 3.81  & 1.14  & 3.60  & 1.12  & 1.41  & 1.22  & 0.58  & 1.06  & -2.52 & 0.76  & -3.37 & 0.71 \\
    C\_BQMall & 87.82 & 0.43  & 15.87 & 1.07  & 18.38 & 0.65  & 13.52 & 1.04  & 12.55 & 0.60  & 8.55  & 0.67  & 4.74  & 1.23 \\
    C\_PartyScene & 102.03 & 0.23  & 10.89 & 0.54  & 13.59 & 0.54  & 4.77  & 0.50  & 3.94  & 0.29  & -0.64 & 0.42  & 0.04  & 0.51 \\
    C\_Racehorses & 48.13 & 1.03  & 14.23 & 0.05  & 16.46 & 0.09  & 7.12  & 0.06  & 9.61  & 0.05  & 4.15  & 1.22  & 2.67  & 0.01 \\
    D\_BasketballPass & 27.64 & 0.42  & 6.75  & 0.72  & 9.91  & 1.01  & 3.82  & 0.73  & 4.75  & 1.05  & 2.74  & 0.70  & 0.16  & 0.56 \\
    D\_BlowingBubbles & 70.31 & 0.73  & 17.77 & 1.73  & 25.97 & 0.75  & 10.95 & 1.11  & 15.25 & 0.61  & 9.27  & 1.16  & 2.45  & 1.54 \\
    D\_BQSquare & 91.74 & 0.42  & 7.22  & 1.05  & 8.99  & 1.01  & 5.01  & 1.28  & 5.88  & 1.03  & -1.02 & 2.54  & 0.30  & 0.96 \\
    D\_Racehorses & 44.49 & 1.13  & 10.13 & 0.27  & 12.56 & 0.20  & 3.40  & 0.34  & 4.44  & 0.24  & -2.76 & 2.00  & -0.56 & 0.20 \\
    \midrule
    Average   & 65.76 & 0.74  & 13.66 & 0.81  & 15.65 & 0.68  & 8.03  & 0.79  & 9.76  & 0.69  & 4.92  & 1.25  & 3.67  & 0.92 \\
    \midrule
    Average Without * & 57.24 & 0.74  & 10.62 & 0.68  & 13.19 & 0.53  & 5.74  & 0.64  & 7.48  & 0.50  & 2.50  & 1.18  & 1.05  & 0.73 \\
    \bottomrule
    \end{tabular}%
  \vspace{0.3cm}
  \caption{RD Performance for LDP Configuration, CQP as Anchor}
  \label{tab:ldp}%
  \begin{tabular}{lcccccccccccccc}
    \toprule
    \multirow{2}[4]{*}{Clip} & \multicolumn{2}{c}{\cite{choi2012rate}} & \multicolumn{2}{c}{\cite{li2014lambda}-Frame} & \multicolumn{2}{c}{\cite{li2014lambda}-LCU} & \multicolumn{2}{c}{\cite{li2018lambda}-Frame} & \multicolumn{2}{c}{\cite{li2018lambda}-LCU} & \multicolumn{2}{c}{\cite{li2017optimal}} & \multicolumn{2}{c}{Proposed} \\
\cmidrule{2-15}          & BDBR  & $\Delta$R    & BDBR  & $\Delta$R    & BDBR  & $\Delta$R    & BDBR  & $\Delta$R    & BDBR  & $\Delta$R    & BDBR  & $\Delta$R    & BDBR  & $\Delta$R \\
    \midrule
    B\_BasketballDrive & 17.70 & 0.54  & 0.54  & 0.64  & 7.37  & 0.61  & 0.53  & 0.64  & 4.51  & 0.54  & 4.44  & 0.62  & 3.60  & 0.64 \\
    B\_BQTerrace & 53.03 & 0.79  & 0.38  & 1.27  & 1.09  & 1.15  & 0.32  & 1.25  & 5.84  & 1.05  & 0.70  & 1.10  & 0.58  & 1.21 \\
    B\_Cactus & 42.82 & 0.11  & -5.61 & 0.10  & -4.42 & 0.03  & -5.56 & 0.04  & -2.13 & 0.02  & -5.23 & 0.02  & -8.29 & 0.04 \\
    B\_Kimono & 33.48 & 0.30  & 8.65  & 0.51  & 14.10 & 0.03  & 6.96  & 1.72  & 9.21  & 0.03  & 6.14  & 0.03  & 6.51  & 0.81 \\
    B\_ParkScene & 39.82 & 0.43  & 1.25  & 0.23  & 4.82  & 0.04  & 0.63  & 0.78  & 2.37  & 0.04  & 0.66  & 0.02  & 1.00  & 0.57 \\
    C\_BasketballDrill & 11.51 & 1.32  & -5.43 & 1.32  & -3.95 & 1.21  & -5.39 & 1.35  & -5.78 & 1.24  & -6.13 & 1.29  & -7.53 & 1.35 \\
    C\_BQMall & 22.38 & 1.18  & 7.08  & 1.09  & 9.08  & 1.02  & 7.08  & 1.09  & 7.48  & 1.04  & 6.51  & 0.99  & 5.74  & 1.06 \\
    C\_PartyScene & 69.89 & 1.51  & 2.16  & 0.69  & 7.43  & 0.70  & 2.17  & 0.69  & 2.77  & 0.59  & 1.32  & 0.62  & 2.45  & 0.66 \\
    C\_Racehorses & 20.67 & 0.07  & 7.63  & 0.29  & 14.84 & 0.57  & 7.79  & 0.11  & 9.94  & 0.07  & 9.38  & 0.07  & 8.31  & 0.06 \\
    C\_BasketballPass & 9.85  & 1.72  & 1.33  & 1.05  & 7.94  & 1.05  & 1.39  & 1.04  & 2.20  & 1.00  & 2.21  & 1.30  & 2.38  & 1.09 \\
    D\_BlowingBubbles & 21.22 & 1.13  & 1.09  & 1.05  & 7.62  & 1.02  & 1.08  & 1.04  & 2.28  & 0.94  & 0.57  & 0.87  & 2.53  & 1.00 \\
    D\_BQSquare & 41.05 & 1.87  & 1.26  & 1.53  & 7.42  & 1.42  & 1.26  & 1.57  & 2.13  & 1.42  & 1.43  & 1.40  & -0.35 & 1.50 \\
    D\_Racehorses & 10.25 & 0.13  & 2.69  & 0.29  & 7.73  & 0.26  & 2.75  & 0.21  & 3.48  & 0.18  & 3.12  & 0.18  & 3.14  & 0.18 \\
    E\_FourPeople & 19.18 & 0.58  & 6.28  & 0.08  & 11.87 & 0.06  & 6.18  & 0.18  & 2.25  & 0.08  & -2.85 & 0.05  & -8.10 & 0.16 \\
    E\_Johnny & 70.91 & 0.47  & 12.79 & 0.15  & 27.62 & 0.06  & 12.80 & 0.12  & 6.67  & 0.06  & 1.83  & 0.05  & -5.39 & 0.09 \\
    E\_Kristen\&Sara & 37.86 & 0.20  & 5.81  & 0.08  & 15.48 & 0.07  & 5.80  & 0.09  & 0.31  & 0.10  & -6.28 & 0.11  & -11.33 & 0.09 \\
    \midrule
    Average & 32.60 & 0.77  & 2.99  & 0.65  & 8.50  & 0.58  & 2.86  & 0.74  & 3.35  & 0.52  & 1.11  & 0.54  & -0.30 & 0.66 \\
    \bottomrule
    \end{tabular}%
    \vspace{0.3cm}
  \caption{RD Performance for LDB Configuration, CQP as Anchor}
  \label{tab:ldb}%
  \begin{tabular}{lcccccccccccccc}
    \toprule
    \multirow{2}[4]{*}{Clip} & \multicolumn{2}{c}{\cite{choi2012rate}} & \multicolumn{2}{c}{\cite{li2014lambda}-Frame} & \multicolumn{2}{c}{\cite{li2014lambda}-LCU} & \multicolumn{2}{c}{\cite{li2018lambda}-Frame} & \multicolumn{2}{c}{\cite{li2018lambda}-LCU} & \multicolumn{2}{c}{\cite{li2017optimal}} & \multicolumn{2}{c}{Proposed} \\
\cmidrule{2-15}          & BDBR  & $\Delta$R    & BDBR  & $\Delta$R    & BDBR  & $\Delta$R    & BDBR  & $\Delta$R    & BDBR  & $\Delta$R    & BDBR  & $\Delta$R    & BDBR  & $\Delta$R \\
    \midrule
    B\_BasketballDrive & 19.04 & 0.61  & 5.62  & 0.69  & 8.64  & 0.67  & 0.94  & 0.69  & 4.93  & 0.58  & 5.01  & 0.66  & 3.58  & 0.68 \\
    B\_BQTerrace & 76.62 & 0.77  & 1.71  & 1.48  & 4.69  & 1.35  & 3.38  & 1.48  & 8.64  & 1.25  & 2.78  & 1.29  & 1.38  & 1.43 \\
    B\_Cactus & 46.83 & 0.13  & -4.81 & 0.10  & -3.52 & 0.03  & -4.16 & 0.05  & -1.03 & 0.03  & -4.22 & 0.02  & -8.26 & 0.04 \\
    B\_Kimono & 35.64 & 0.28  & 11.81 & 0.55  & 13.58 & 0.03  & 6.48  & 1.71  & 8.52  & 0.02  & 5.57  & 0.02  & 5.74  & 0.84 \\
    B\_ParkScene & 41.17 & 0.42  & 2.38  & 0.32  & 4.93  & 0.06  & 0.95  & 0.77  & 2.76  & 0.05  & 0.94  & 0.04  & 1.10  & 0.64 \\
    C\_BasketballDrill & 38.42 & 1.67  & -6.33 & 1.33  & -3.32 & 1.25  & -4.77 & 1.40  & -5.39 & 1.25  & -5.56 & 1.29  & -7.00 & 1.37 \\
    C\_BQMall & 22.42 & 1.15  & 9.32  & 1.10  & 9.54  & 1.06  & 7.86  & 1.12  & 7.81  & 1.07  & 6.56  & 1.02  & 6.18  & 1.08 \\
    C\_PartyScene & 58.28 & 1.21  & 4.89  & 0.70  & 8.13  & 0.70  & 2.40  & 0.69  & 3.05  & 0.60  & 1.56  & 0.63  & 1.71  & 0.68 \\
    C\_Racehorses & 20.98 & 0.04  & 13.31 & 0.26  & 15.53 & 0.51  & 8.04  & 0.06  & 10.63 & 0.03  & 10.32 & 0.02  & 8.39  & 0.05 \\
    C\_BasketballPass & 8.72  & 1.63  & 5.65  & 1.09  & 8.22  & 1.11  & 1.63  & 1.12  & 2.17  & 1.03  & 2.49  & 1.35  & 2.55  & 1.12 \\
    D\_BlowingBubbles & 19.75 & 1.18  & 4.94  & 1.05  & 7.66  & 1.03  & 0.91  & 1.02  & 2.00  & 0.93  & 0.02  & 0.87  & 2.19  & 1.03 \\
    D\_BQSquare & 41.26 & 1.67  & 8.06  & 1.56  & 10.93 & 1.48  & 2.85  & 1.62  & 3.66  & 1.42  & 3.02  & 1.45  & 0.61  & 1.47 \\
    D\_Racehorses & 10.30 & 0.13  & 7.03  & 0.28  & 8.63  & 0.18  & 3.15  & 0.14  & 3.62  & 0.12  & 3.34  & 0.12  & 3.30  & 0.14 \\
    E\_FourPeople & 19.21 & 0.71  & 14.76 & 0.10  & 12.53 & 0.07  & 6.77  & 0.20  & 3.07  & 0.08  & -2.21 & 0.07  & -7.44 & 0.19 \\
    E\_Johnny & 69.72 & 0.71  & 34.84 & 0.37  & 30.70 & 0.26  & 16.58 & 0.34  & 9.64  & 0.26  & 2.95  & 0.26  & -2.68 & 0.31 \\
    E\_Kristen\&Sara & 39.79 & 0.22  & 18.94 & 0.08  & 15.94 & 0.10  & 7.26  & 0.08  & 1.55  & 0.10  & -5.16 & 0.10  & -10.30 & 0.09 \\
    \midrule
    Average & 35.51 & 0.78  & 8.26  & 0.69  & 9.55  & 0.62  & 3.77  & 0.78  & 4.10  & 0.55  & 1.71  & 0.57  & 0.07  & 0.70 \\
    \bottomrule
    \end{tabular}%
\end{table*}%

\begin{figure*}[htbp]
\setcounter{subfigure}{0}
\subfigure[PeopleOnStreet Bits]{\includegraphics[width=0.48\textwidth, clip=true]{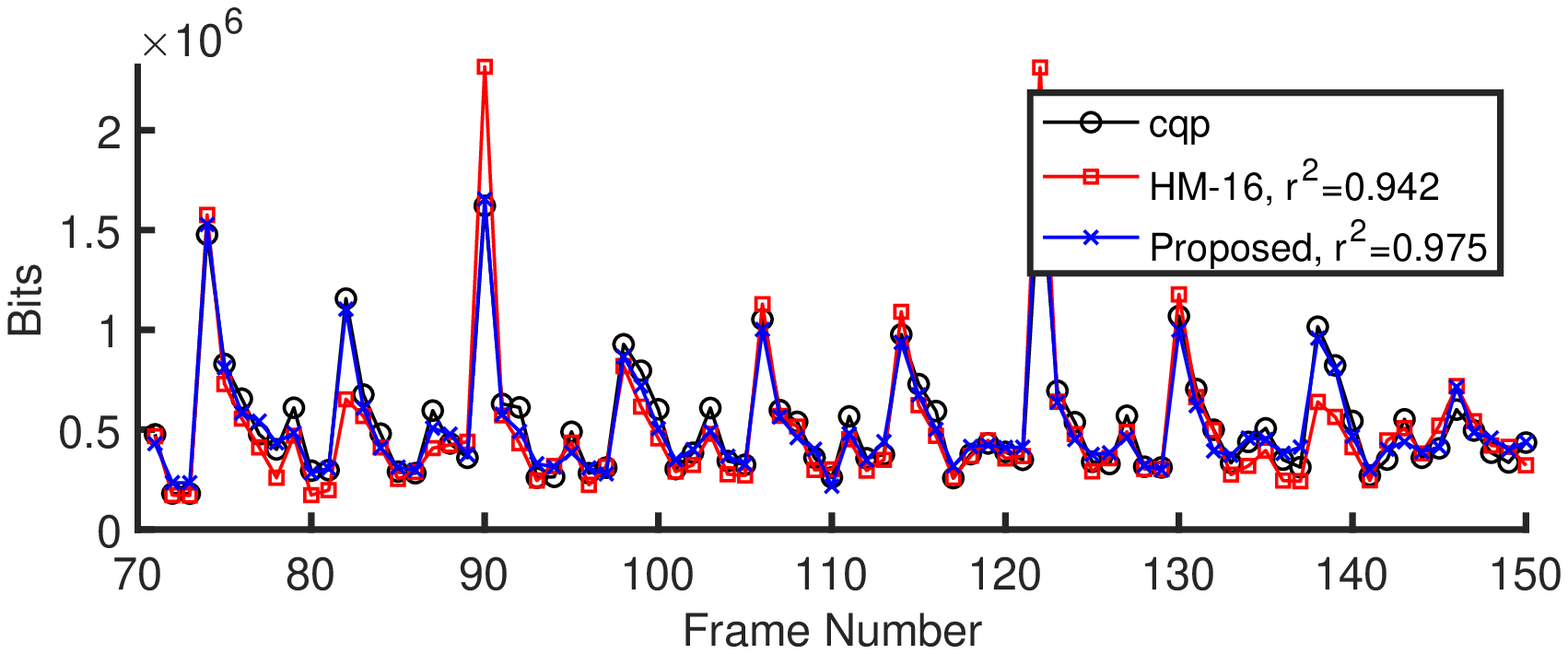}}
\subfigure[PeopleOnStreet PSNR]{\includegraphics[width=0.48\textwidth, clip=true]{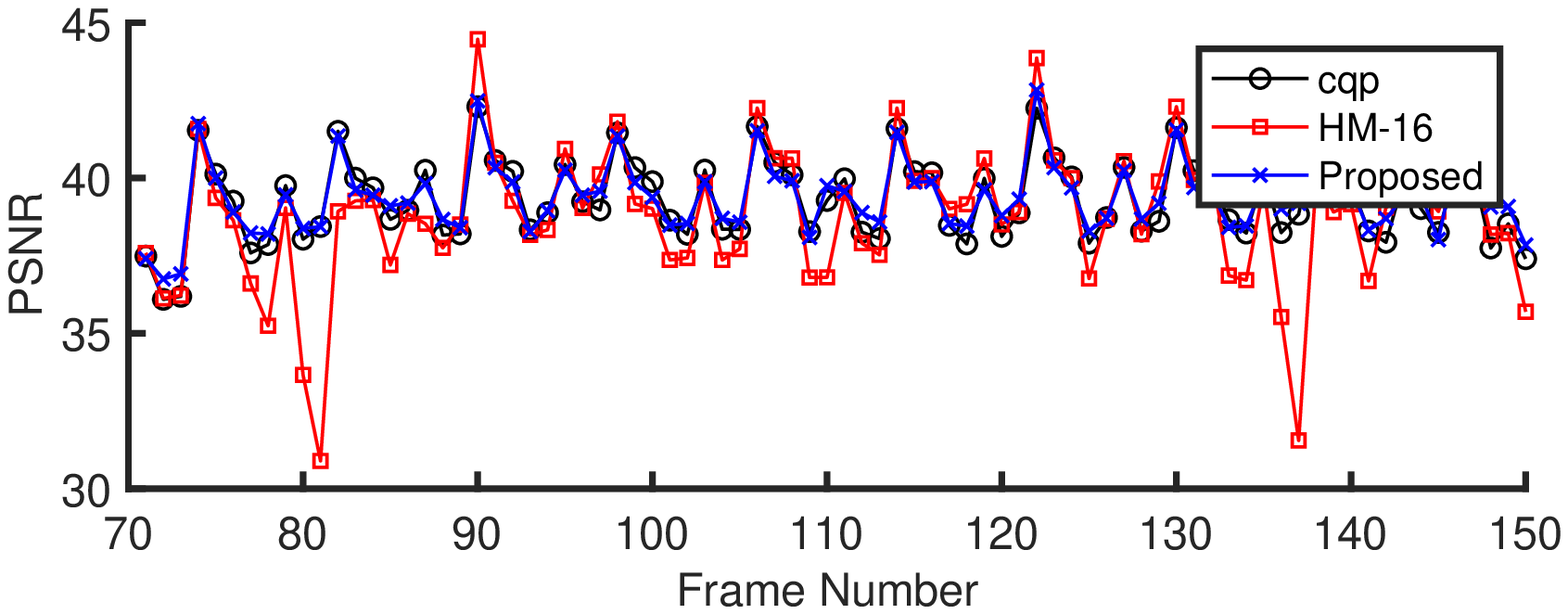}}\\
\subfigure[BlowingBubbles Bits]{\includegraphics[width=0.48\textwidth, clip=true]{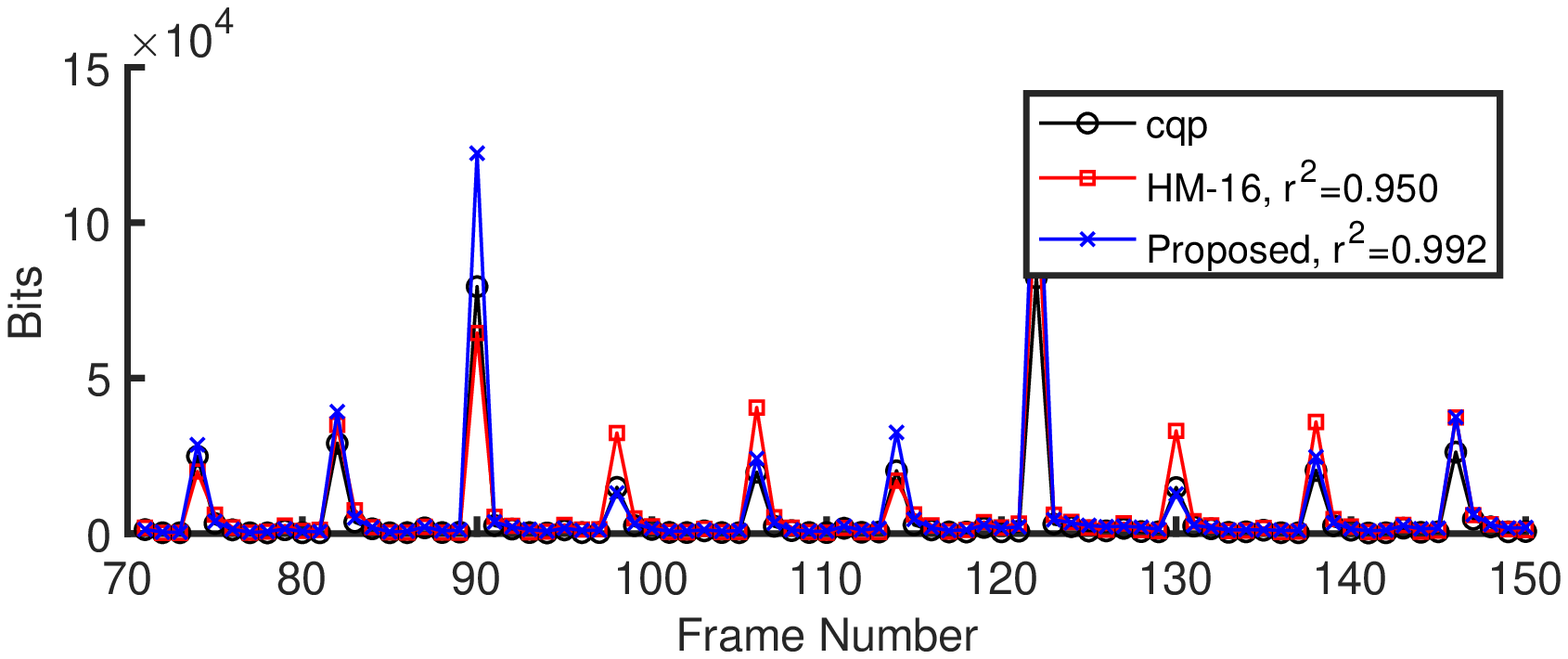}}
\subfigure[BlowingBubbles PSNR]{\includegraphics[width=0.48\textwidth, clip=true]{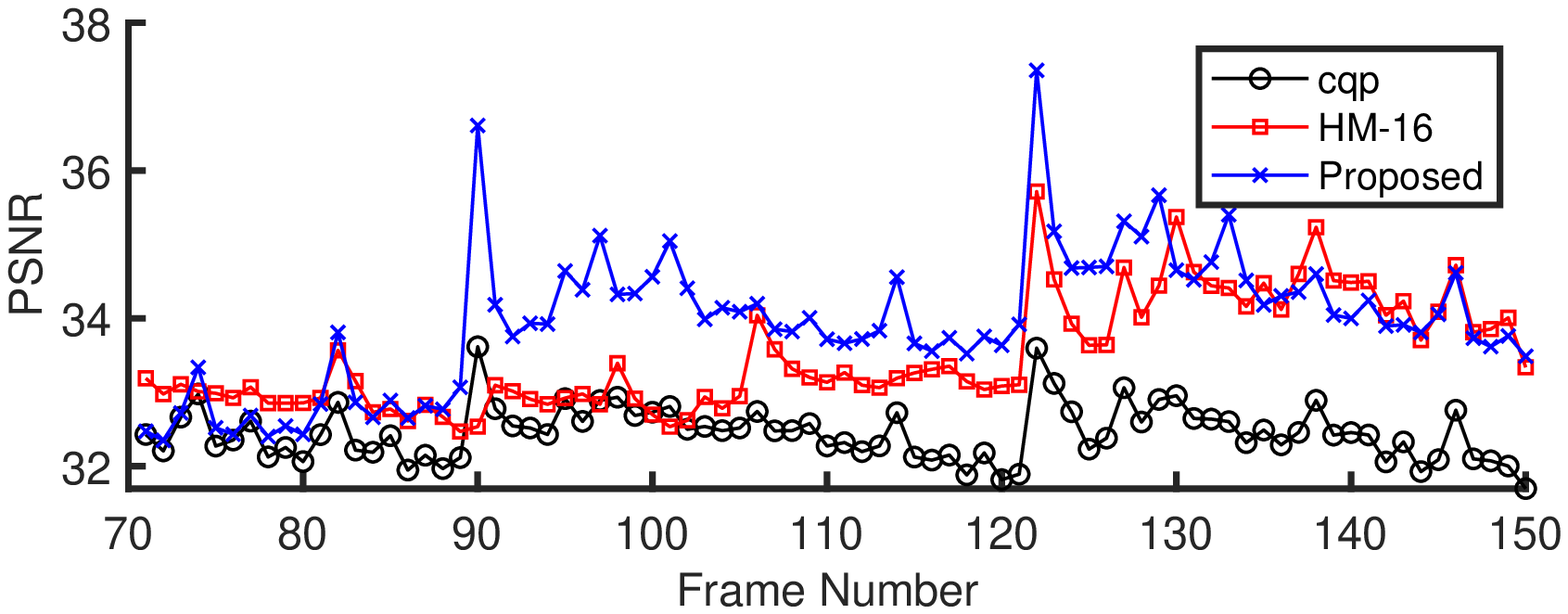}}
\caption{Per-Frame Bits and PSNR for PeopleOnStreet and BlowingBubbles}
\label{fig:bitspsnr}

\setcounter{subfigure}{0}
\subfigure[RA, PartyScene, 480p]{\includegraphics[width=0.48\textwidth, clip=true]{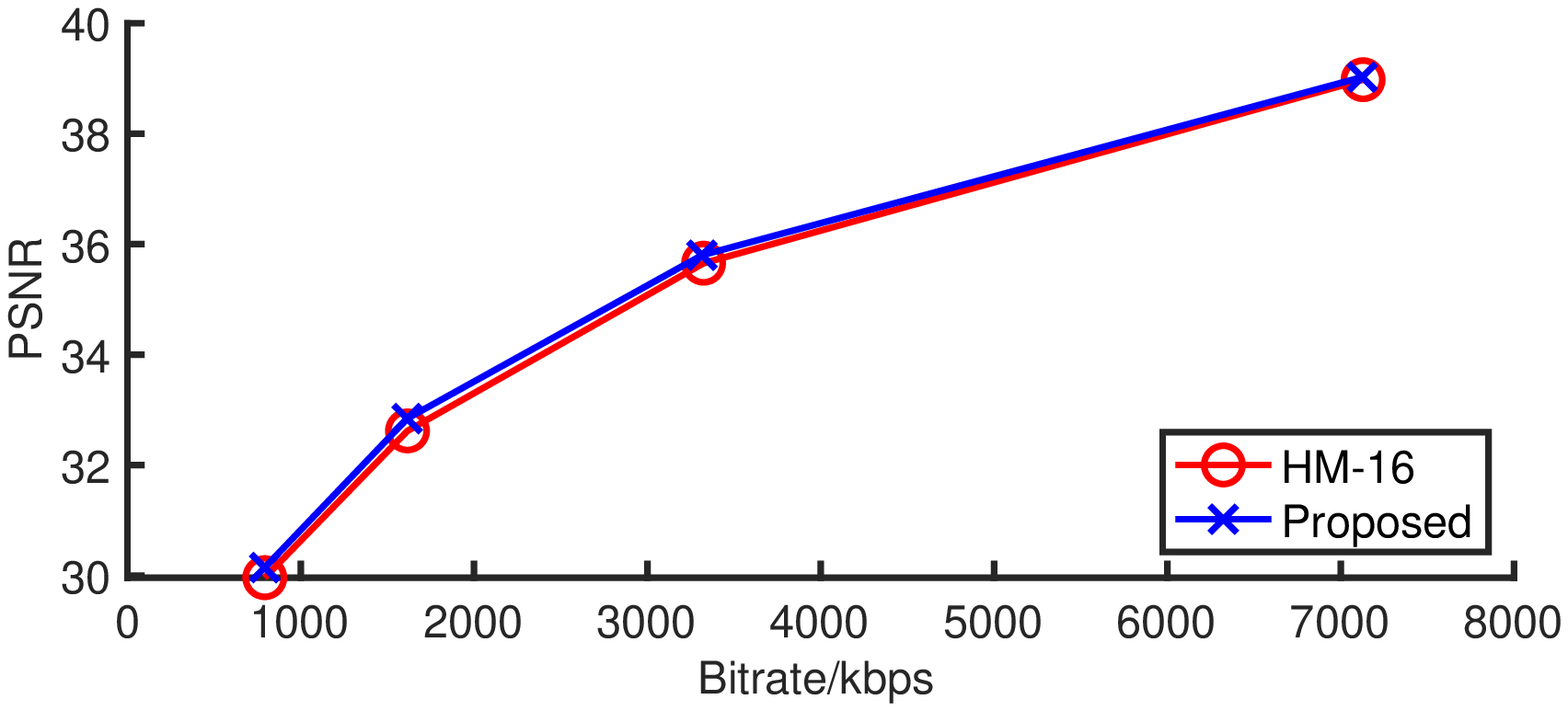}}
\subfigure[RA, BlowingBubbles, 240p]{\includegraphics[width=0.48\textwidth, clip=true]{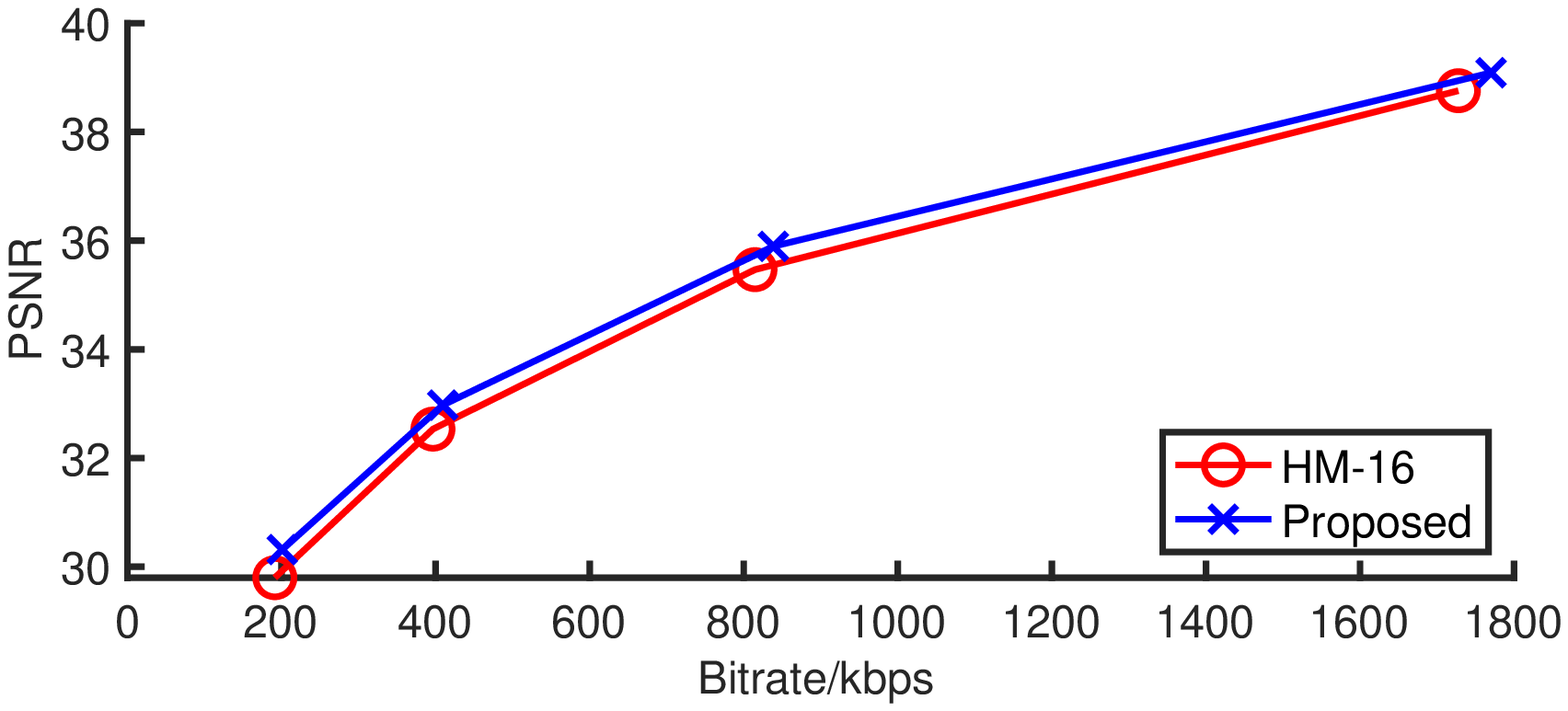}}\\
\subfigure[LDP, BQTerrace, 1080p]{\includegraphics[width=0.48\textwidth, clip=true]{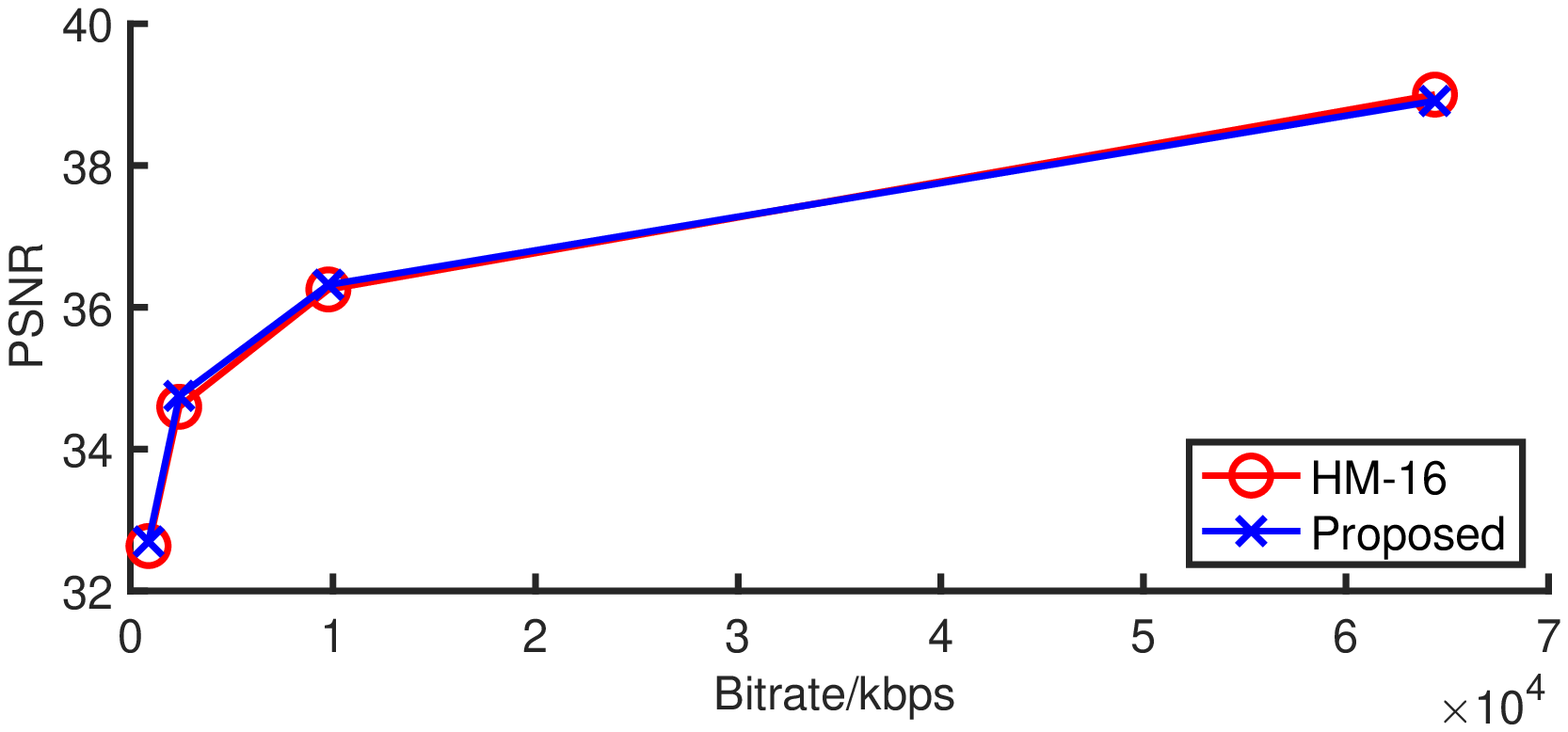}}
\subfigure[LDP, FourPeople, 720p]{\includegraphics[width=0.48\textwidth, clip=true]{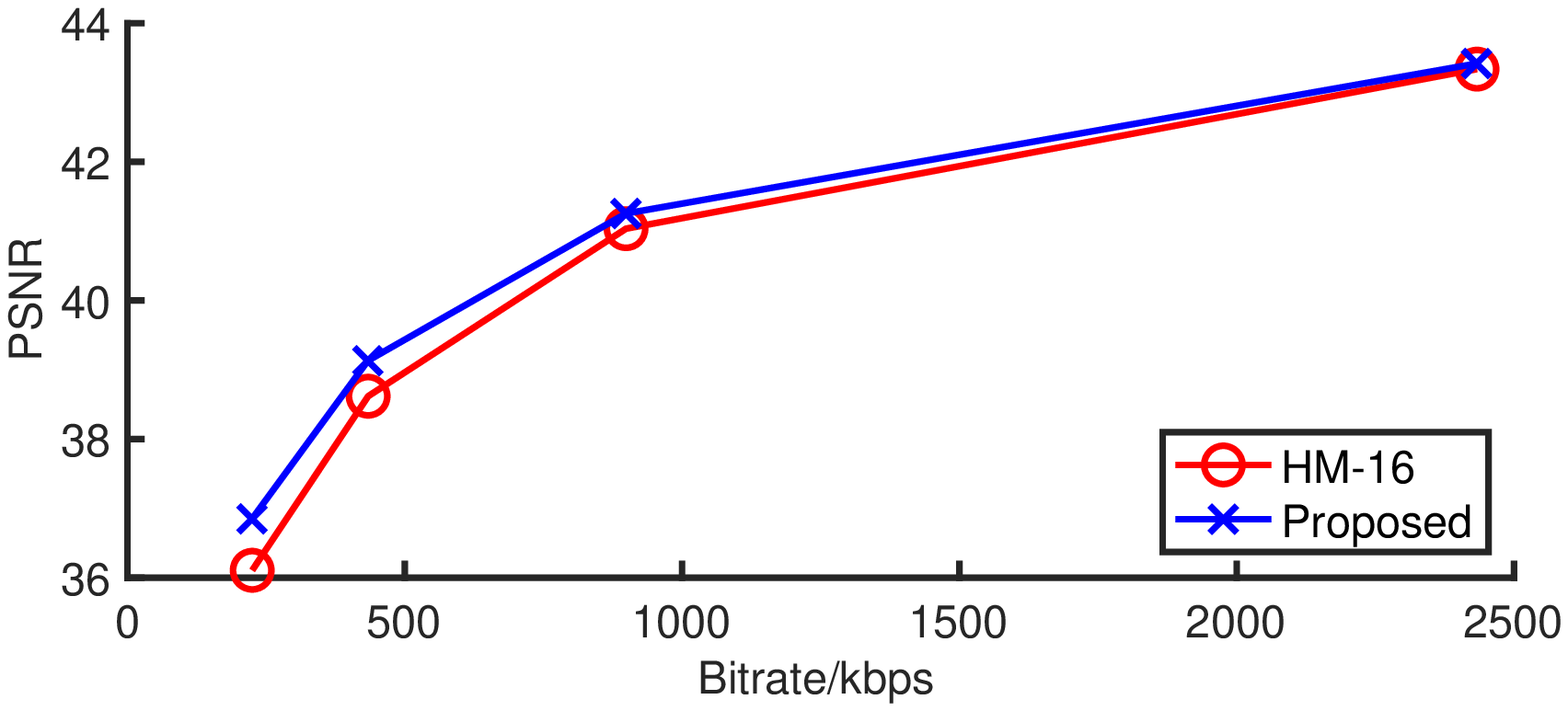}}\\
\subfigure[LDB, Cactus, 1080p]{\includegraphics[width=0.48\textwidth, clip=true]{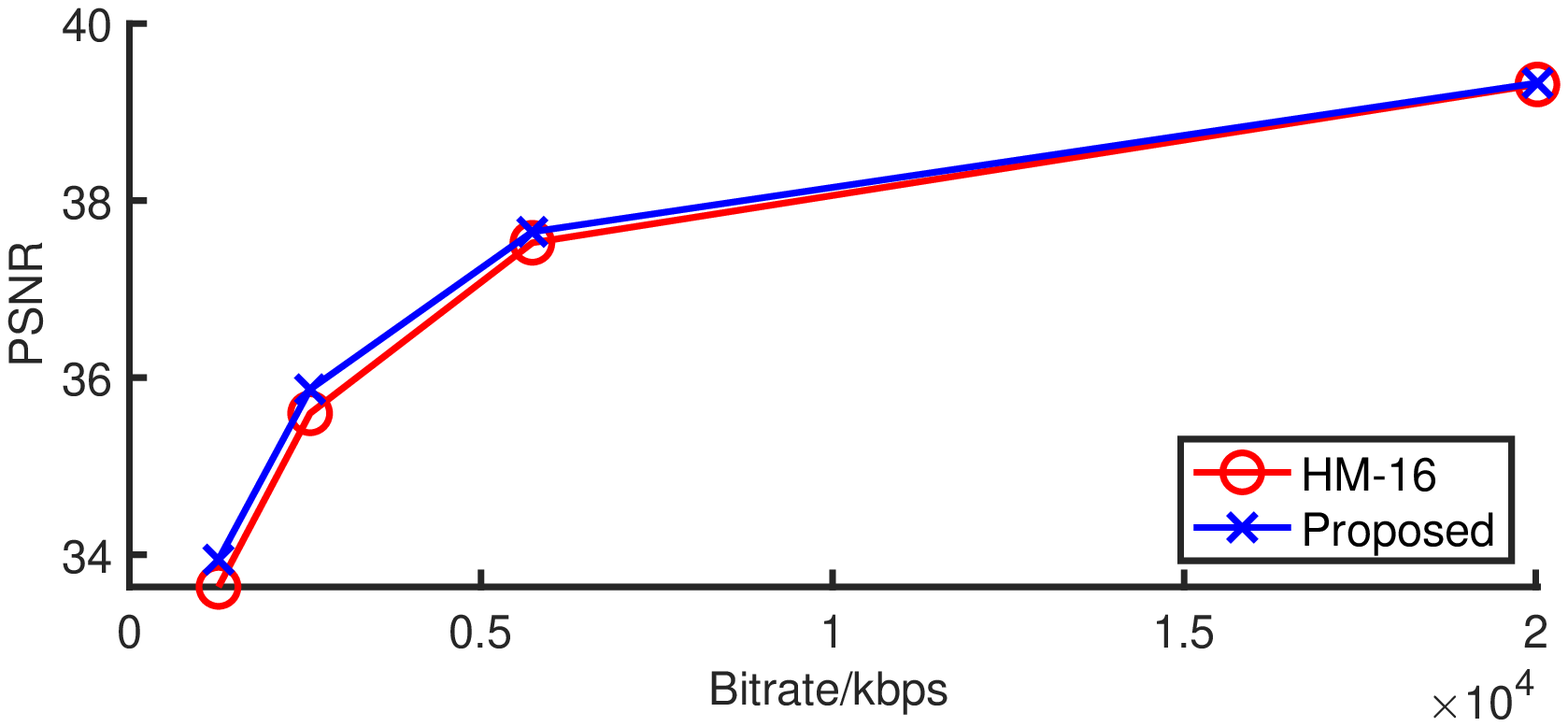}}
\subfigure[LDB, Johnny, 720p]{\includegraphics[width=0.48\textwidth, clip=true]{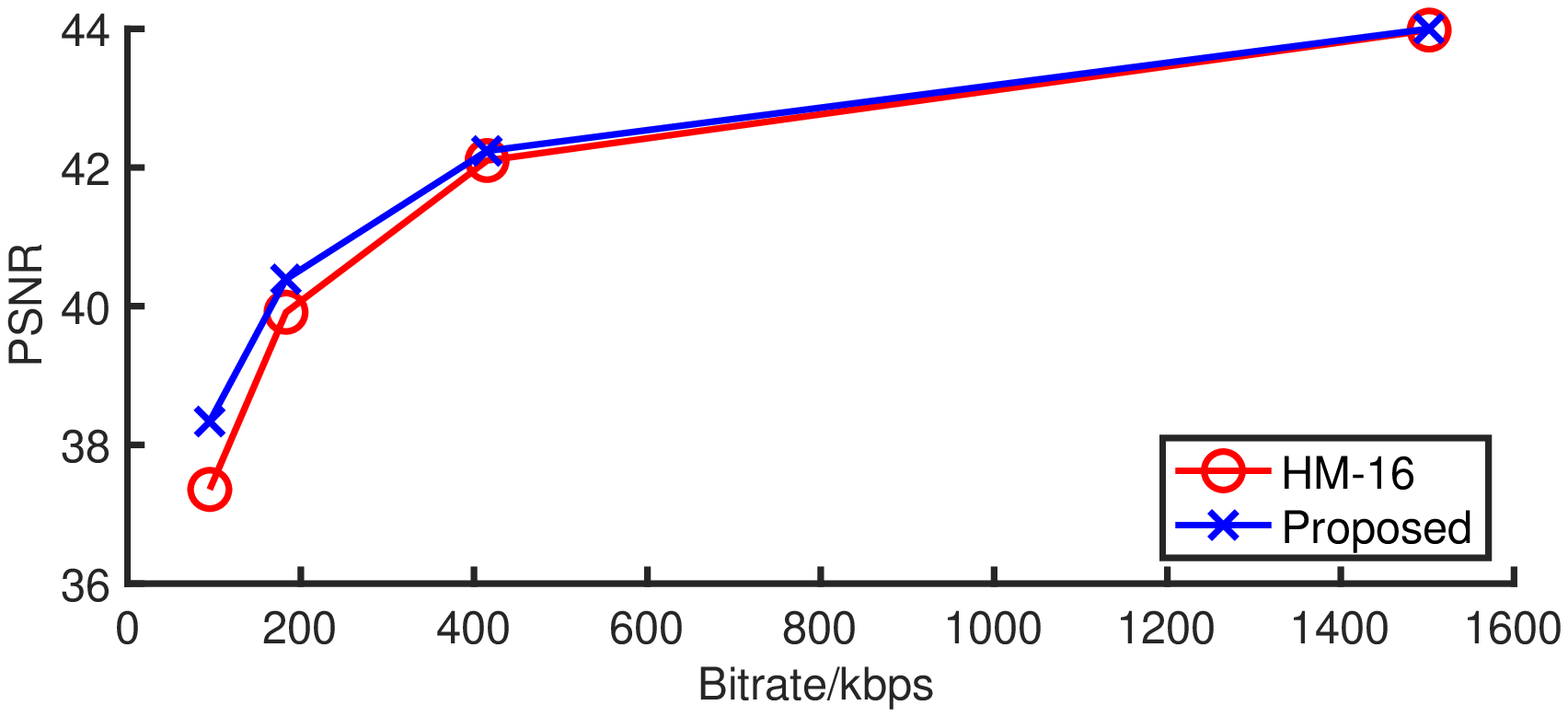}}
\caption{Comparisons of the RD Curves for Different Sequences}
\label{fig:rdcurve}
\end{figure*}

\section{Experimental Results}
\label{sec:exp}

\subsection{Experiments Set-Up}
To evaluate the performance of the proposed algorithm, the 
proposed algorithm was implemented in the HEVC reference 
software HM-16.9\footnote{HM-16.x adopts the same RC algorithm \cite{li2018lambda}
as that of HM-14 as well as the same coding syntaxes. It is widely 
acknowledged that there is no gain in coding efficiency introduced since HM-14.}.
Besides the proposed algorithm, the state-of-the-art rate control algorithms 
\cite{choi2012rate, li2014lambda, li2018lambda, 
li2017optimal, he2017efficient} were also tested in the experiment.
It should be noted that the algorithms in \cite{li2014lambda} and 
\cite{li2018lambda} provide two modes each, allowing LCU-level 
separate model or not. The corresponding results of two modes are denoted with suffix
Frame/LCU in the tables. The default setting of HM enables 
LCU-level separate model. It is found that the rate control algorithm in HM
is able to achieve a higher coding efficiency and a higher 
bit rate error without LCU-level separate model, so the two 
modes were both tested in the experiments.

The HEVC common test condition \cite{bossen2013common}\footnote{More 
details of the HEVC common test condition can be found in\\
\url{http://phenix.int-evry.fr/jct}} is used as the core test 
of the proposed experiment. All 20 videos (roughly 10 seconds each) 
in class A to E of HEVC common test condition are used as the test 
sequences to cover videos of various video characteristics.
According to the HEVC common test condition, the sequences in 
class B, C, D and E are tested for the LD test, including low delay P 
(LDP) configuration and low delay B (LDB) configuration,
while the sequences in classes A, B, C and D are tested in the RA test. 
As instructed in the common test condition, the test sequences were 
first encoded by HM-16.9 using the 
CQP mode (QP in \{22,27,32,37\}) and three configurations to 
generate the target average bit rates of the later rate control (ABR)
encoding. The results of the CQP mode are also used as the 
anchor of the comparison, which is widely accepted as 
the statistical upper limit of the coding efficiency that one-pass rate 
control algorithms can approach.

Then different rate control algorithms were used to encode 
the test sequences into the corresponding target bit rates generated by the CQP mode.
As to the state-of-the-art algorithms, it should be noted that the algorithm in \cite{li2017optimal},
the best rate control algorithm for LDP and LDB configurations,
was designed dedicated for LD, while the 
algorithm in \cite{he2017efficient}, so far the best RC algorithm 
for RA, was solely proposed for RA. As a result, the algorithms 
in \cite{li2017optimal} and \cite{he2017efficient} were only 
tested for LD and RA respectively.

As instructed in the HEVC common test condition, the algorithms were 
evaluated using two metrics, PSNR based coding efficiency and rate control accuracy in sequence level.
Four points of encoding results (bit rate and MSE based YUV PSNR)
were used to calculate the BDBR between the CQP mode and 
RC algorithms. BDBR was proposed in 
\cite{bjontegaard2008improvements}, which estimates the
delta bit rate that is needed for the current codec 
to achieve a same quality as compared with the 
anchor codec. A negative value of BDBR suggests 
an average bit rate saving for a given quality after 
compression, i.e. gain in coding efficiency. Rate control accuracy was measured by the
averaged absolute bit rate error per sequence using the following formula:
\begin{eqnarray}
\Delta R = \frac{|R_{out}-R_{target}|}{R_{target}}\times 100,
\end{eqnarray}
where $|.|$ operator is to get the absolute value. The 
experiments were conducted on a server with dual Intel Xeon 
CPU E5-2695 v2 without optimization on parallelism and SIMD. 

To better evaluate the performance of the proposed algorithm, 
subjective quality (SSIM\cite{wang2004image}) based coding 
efficiency (BDBR-SSIM) is also analyzed in the proposed experiment 
besides the HEVC common test condition which only compares the 
objective quality (PSNR) based coding efficiency.

In the subjective experiment, BDBR-SSIM results 
are used to evaluate the performance of the proposed algorithm.
SSIM value is designed to lie in the range of [0, 1], a greater SSIM 
value suggesting a better subjective quality. However, SSIM is not a 
distance metric, which is usually first translated into a dB value 
(${SSIM}_{dB}$) using Equation (\ref{eqn:ssim}) for a better fitting 
in the BDBR tool, especially when SSIM values are very close to 1.
\begin{eqnarray}
\label{eqn:ssim}
{SSIM}_{dB} = min(-10 * {log}_{10}(1 - ssim), 100),
\end{eqnarray}

In addition, the respective impact caused by the modules in the 
proposed algorithm are also tested and discussed, including 
objective and subjective quality based coding efficiency as well as computing complexity.

\subsection{Performance for HEVC Common Test Condition}
\label{subsec:performance}

Table \ref{tab:ra} gives the results of various 
algorithms for RA. The default algorithm 
in HM (\cite{li2018lambda}-LCU) is 9.76\% worse than the CQP 
mode, while the state-of-the-art RC algorithm for RA 
\cite{he2017efficient} reduces the gap to 4.92\% at a cost of 
nearly doubled rate control error.
The proposed algorithm is only 3.67\% away from CQP,
while the rate control error is 0.92\%, which is much lower than \cite{he2017efficient}.
It should be noted that among the test sequences, SteamLoco 
(marked with $*$ in Table \ref{tab:ra}) is an outlier due to 
its extremely high complexity. SteamLoco is a video captured by a 
moving camera, which contains a running steam locomotive with 
plenty of billowing steam, i.e. rich and volatile texture with 
limited temporal similarity. Therefore, SteamLoco is not 
suitable for hierarchical schemes which were built based on the 
assumption of high temporal similarity. None of the algorithms 
is able to properly handle this video, and the
BDBR losses of various RC algorithms over CQP are all around 45\%.
The BDBR results of SteamLoco are much greater than other 
videos, which has a significant influence on the average 
results. Therefore, the bottom row of Table \ref{tab:ra} also 
provides the averaged results without SteamLoco to better 
compare the performance on other videos. Under that criterion, 
the proposed algorithm is only 1.05\% worse than the CQP mode, 
while the algorithms in \cite{li2018lambda}-LCU and 
\cite{he2017efficient} are 5.74\% and 2.50\% worse than the CQP mode.

Table \ref{tab:ldp} and \ref{tab:ldb} give the 
results of the LDP and LDB configurations. 
Experimental results show that the proposed algorithm is on average
0.30\% better than CQP for LDP and only 0.07\% worse than CQP 
for LDB with a very low bit rate error.
The proposed algorithm is the first rate control algorithm that is able to reach the statistically
upper limit of the coding efficiency of single-pass rate 
control algorithms for LD. On the other side, 
\cite{li2018lambda}-LCU, the default RC algorithm in HM, is 3.35\% and 4.10\% worse 
than CQP for LDP and LDB, while the best RC 
algorithm for LD in \cite{li2017optimal} is 1.11\% and 1.71\% 
worse than CQP for LDP and LDB respectively.
It should be noted that the results of \cite{li2017optimal} 
is slightly different from the results in their paper,
because a different setting of option 
``KeepHierarchicalBit'' was used in \cite{li2017optimal}.
In the experiment of \cite{li2017optimal}, option ``KeepHierarchicalBit'' was 
set to 1 for the anchor configuration, while the default 
setting in HM-16.9 is 2, which outperforms 1 in coding efficiency.

It should be noted that the proposed algorithm achieves a better 
coding efficiency than the CQP mode for some videos.
This result is not against the claim that the coding 
efficiency of the CQP mode is the ``statistical'' upper limit of single-pass RC algorithms.
The coefficients in the QP-$\lambda$ relationship in Equation 
(\ref{eqn:qp}) and the hierarchical structures are statistically 
optimal for the HEVC common test condition, which are not 
necessarily optimal for any single video.
As a result, it is possible that a RC algorithm produces a rate 
allocation and coding parameters selection scheme that is closer 
to the per-sequence optimum than the CQP mode.

Fig. \ref{fig:bitspsnr} gives some examples of per-frame
rate consumption and PSNR using the proposed algorithm, 
the default RC algorithm in HM \cite{li2018lambda} and the CQP mode, 
where $r^2$ is the correlation coefficient between the output 
per-frame rates of RC algorithms and CQP.
The CQP mode is usually considered as the statistical upper bound of 
the coding efficiency of single-pass rate control algorithms, 
because the CQP mode was designed to encode a video into a 
reasonable distribution of quality after compression. The output 
per-frame rates of CQP mode suggest a favorable distribution of
per-frame rates to hit the favorable distribution of quality after compression. As a result, a
higher correlation indicates a better rate allocation.
Fig. \ref{fig:bitspsnr}-(a) and \ref{fig:bitspsnr}-(c) show that the proposed
algorithm is able to produce per-frame rates that are
much closer to the CQP output than \cite{li2018lambda}.
Fig. \ref{fig:bitspsnr}-(b) gives an example of improper 
overflow compensation of \cite{li2018lambda}, which causes a
significant quality deterioration after a frame that consumes too many 
bits. On the contrary, the proposed algorithm encodes that part into a 
more temporally consistent quality. Fig. \ref{fig:rdcurve} 
gives some examples of RD curves comparison for different 
configurations, which show that the proposed algorithm is 
able to steadily improve the coding efficiency for a wide 
range of bit rates. Fig. \ref{fig:rdcurve}-(d) and \ref{fig:rdcurve}-(f) show 
that the proposed algorithm is able to provide a higher gain for 
the cases targeting very low bit rates.

\subsection{Module-level Subjective and Objective Performance}
\label{subsec:module}

The proposed module-level experiment includes objective and 
subjective performance evaluation as well as complexity analysis.
As some modules in the proposed algorithm are designed to 
work with the proposed R-D-$\lambda$ model, like the new update 
scheme, it is hard to test and analyze those modules separately. 
Therefore, the module-level analysis is conducted in an incremental way.
The proposed algorithm is separated into four phases, namely,
\begin{enumerate}
\item \textit{Phase 1}: The proposed R-D-${\lambda}$ model in 
Equation (\ref{eqn:lambdabpp}) is used to replace the model used in 
\cite{li2014lambda} with the QP-$\lambda$ relation in Equation 
(\ref{eqn:newqp}). The model coefficients are updated using the 
update scheme proposed in \cite{li2014lambda}.
\item \textit{Phase 2}: The proposed hierarchical initialization 
scheme is implemented on top of \textit{Phase 1}.
\item \textit{Phase 3}:	 The proposed rate allocation scheme is 
implemented on top of \textit{Phase 2}.
\item \textit{Phase 4}: The proposed update scheme is used to 
replace the update scheme proposed in \cite{li2014lambda} on top of 
\textit{Phase 3}. \textit{Phase 4} is the proposed algorithm.
\end{enumerate}

\begin{table}[b]
  \centering
  \caption{Per-module Analysis of the Proposed Algorithm}
    \begin{tabular}{cccccc}
    \toprule
    \multicolumn{2}{c}{HM-RC\cite{li2018lambda} as Anchor} & \textit{Phase 1} & \textit{Phase 2} & \textit{Phase 3} & \textit{Phase 4} \\
    \midrule
    \multirow{3}[2]{*}{RA} & BDBR-PSNR & -4.35\% & -4.37\% & -4.49\% & -5.44\% \\
          & BDBR-SSIM & -1.26\% & -1.48\% & -1.48\% & -2.16\% \\
          & $\Delta$T & +1.70\% & +1.74\% & +1.70\% & +1.67\% \\
    \midrule
    \multirow{3}[2]{*}{LDP} & BDBR-PSNR & -1.68\% & -1.75\% & -3.27\% & -3.62\% \\
          & BDBR-SSIM & -0.88\% & -1.41\% & -2.76\% & -3.13\% \\
          & $\Delta$T & +1.10\% & +0.69\% & -0.08\% & +0.18\% \\
    \midrule
    \multirow{3}[2]{*}{LDB} & BDBR-PSNR & -1.78\% & -1.79\% & -3.54\% & -3.93\% \\
          & BDBR-SSIM & -0.85\% & -1.34\% & -2.98\% & -3.42\% \\
          & $\Delta$T & +0.99\% & +0.63\% & +0.01\% & -0.40\% \\
    \bottomrule
    \end{tabular}%
  \label{tab:submodule}%
\end{table}%

Table \ref{tab:submodule} shows the performance of each phase of 
the proposed algorithm. The proposed R-D-$\lambda$ model 
(\textit{Phase 1}) provides the majority of the gain in objective 
quality based coding efficiency (BDBR-PSNR), which is 4.35\%/5.44\% 
for RA, 1.68\%/3.62\% for LDP and 1.78\%/3.93\% for LDB.
The gain caused by the proposed model is higher for RA than LD 
because RA allows bi-directional referencing with more available reference 
frames, which greatly increases the coding efficiency of the frames 
of a higher frame level. The gap between the bit rates of lossy 
encoding and lossless encoding is lower for RA than LD, so the 
proposed model brings a higher gain for RA.

In addition, the proposed algorithm results in a similar gain 
in subjective quality based coding efficiency, though the parameters 
in the proposed algorithm was tuned using BDBR-PSNR, which proves 
that the proposed algorithm is effective for improving both the
objective and subjective quality after compression.

Table \ref{tab:submodule} also provides the results on the 
complexity change caused by the proposed algorithm. An increase of 
0.5\% in complexity is observed on average for three configurations, 
which is usually within the range of measurement noise and therefore 
can be considered negligible.

\section{Conclusion and Discussion}
\label{sec:conclusion}
In this paper, we propose a novel generalized R-D-$\lambda$ 
model to better model the relationship between rate, 
distortion and $\lambda$. Based on the new model, a novel average bit rate
control (ABR) scheme for HEVC is designed, which includes
hierarchical initialization, LMS based update for model coefficients as well as 
amortization and smooth window joint rate allocation.
Experimental results of implementing the proposed algorithm into 
the HEVC reference software HM-16.9 shows that the proposed rate 
control algorithm is able to achieve the best coding efficiency 
among the state-of-the-art RC algorithms, which is only 3.67\% 
and 0.07\% worse than CQP for RA and LDB configurations and 
0.3\% better than CQP for LDP, while rate control accuracy 
and encoding speed are hardly impacted.

In future, we will keep investigating the following issues.
The proposed algorithm is designed and tested using the 
HEVC common test condition, which uses a fixed reference structure 
and an ABR scheme. However, real-world applications are usually 
much more complicated than that, which may require CBR, adaptive 
reference structure and some optimizations on specific usages like
screen content videos and videos of ultra high resolutions (4K, 8K 
and even higher). It will be valuable to optimize the 
proposed algorithm for various application scenarios respectively.
In addition, some of the variables, e.g. the values in Equation 
(\ref{eqn:newqp}), are set to fixed numbers in the proposed 
algorithm, which were selected through tuning experiments.
In fact, the selection of those values is a chicken-and-egg
problem. It will be beneficial to explore a way to
interactively determine these values.


\bibliography{newrc}
\bibliographystyle{IEEEtran}

\vspace{-1cm}
\begin{IEEEbiography}[{\includegraphics[width=1in,height=1.25in,clip,keepaspectratio]{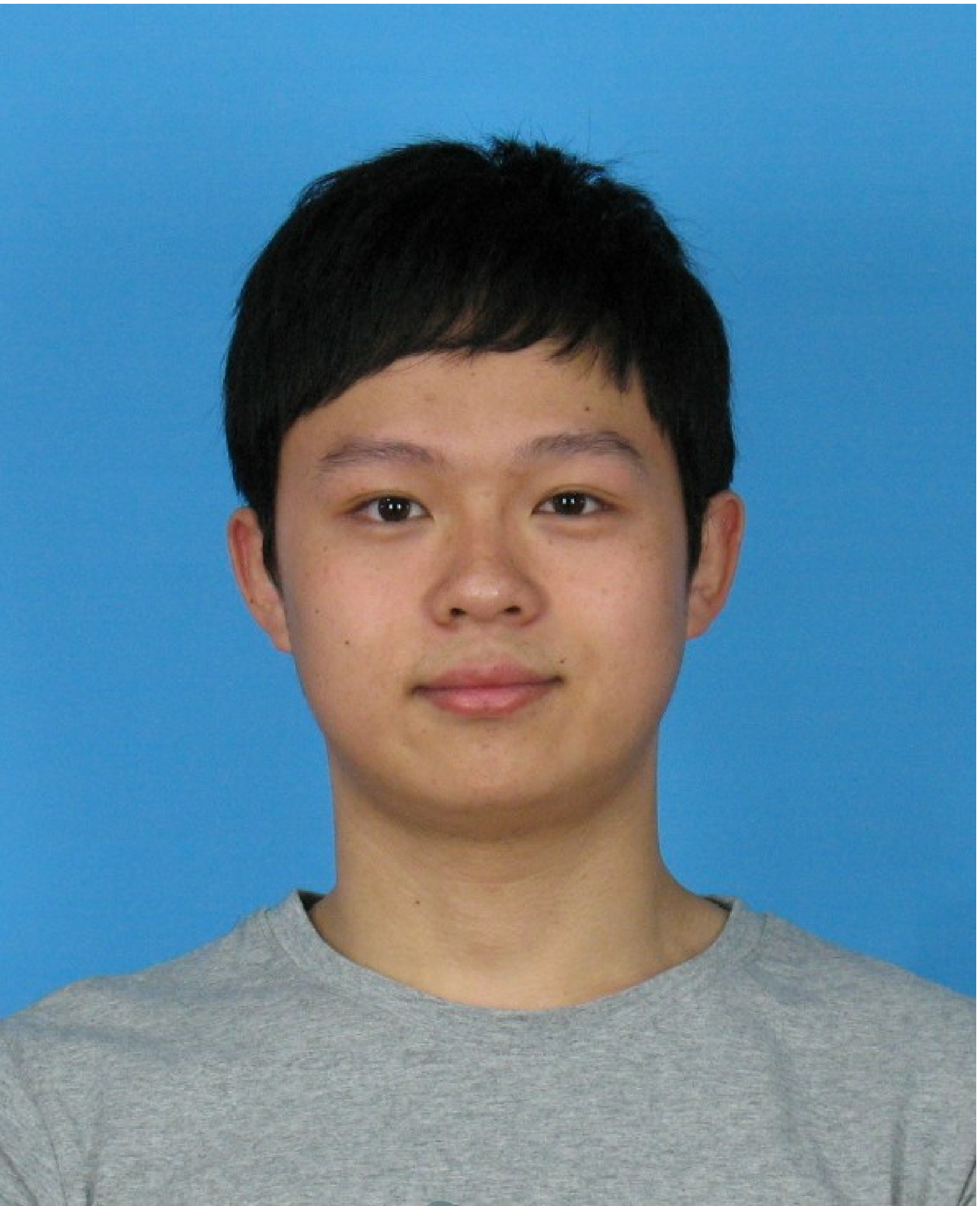}}]{Minhao Tang} received the B.S. degree from the Department of Electronic Engineering, Tsinghua University, Beijing, China, in 2014, and Ph.D degree from the Department of Computer Science and Engineering, Tsinghua University in 2019.
Dr. Tang is a senior researcher at Media Lab of Tencent. \\
\indent Dr. Tang's research interests include video encoding and transcoding and high-quality panoramic videos production. Dr. Tang was a nominee for the 2016 IEEE Trans. CSVT Best Paper Award and 2017 ICME 10K Best Paper Award.
The high quality video encoding/transcoding solution project Dr. Tang developed won 2016 Frost \& Sullivan best practice in Enabling Technology Leadership award.
\end{IEEEbiography}

\vspace{-1cm}



\begin{IEEEbiography}[{\includegraphics[width=1in,height=1.25in,clip,keepaspectratio]{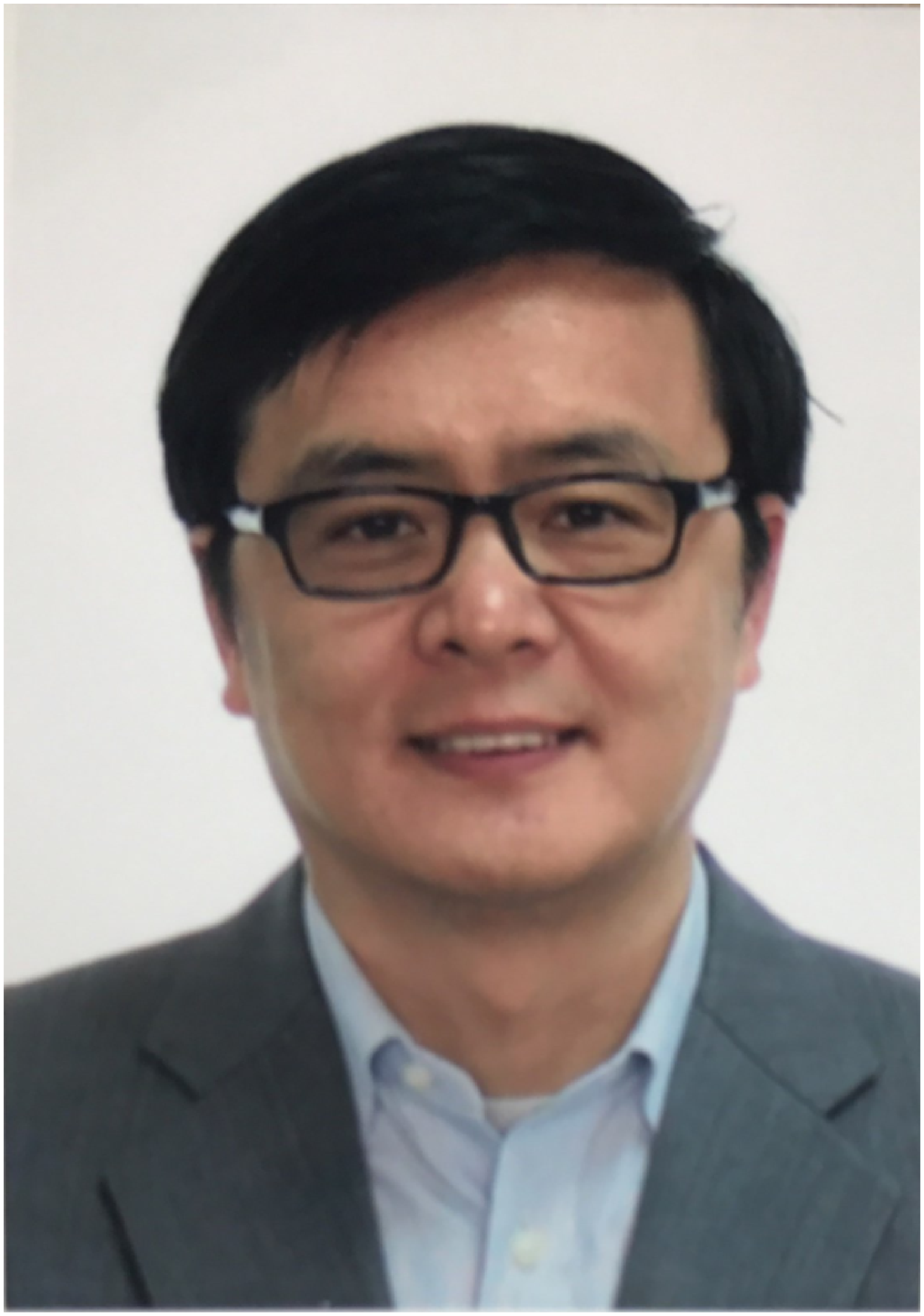}}]{Jiangtao Wen} received the BS, MS and Ph.D. degrees with honors from Tsinghua University, Beijing, China, in 1992, 1994 and 1996 respectively, all in Electrical Engineering.\\
\indent Dr. Wen's research focuses on multimedia communication over challenging networks and computational photography. He has authored many widely referenced papers in related fields. Products deploying technologies that Dr. Wen developed are currently widely used worldwide. Dr. Wen holds over 40 patents with numerous others pending. Dr. Wen is an Associate Editor for IEEE Transactions Circuits and Systems for Video Technologies (CSVT). He is a recipient of the 2010 IEEE Trans. CSVT Best Paper Award and a nominee for the 2016 IEEE Trans CSVT Best Paper Award.
Dr. Wen was elected a Fellow of the IEEE in 2011. He is the Director of the Research Institute of the Internet of Things of Tsinghua University, and a Co-Director of the Ministry of Education Tsinghua-Microsoft Joint Lab of Multimedia and Networking.
Besides teaching and conducting research, Dr. Wen also invests in high technology companies as an angel investor.
\end{IEEEbiography}

\vspace{-1cm}

\begin{IEEEbiography}[{\includegraphics[width=1in,height=1.25in,clip,keepaspectratio]{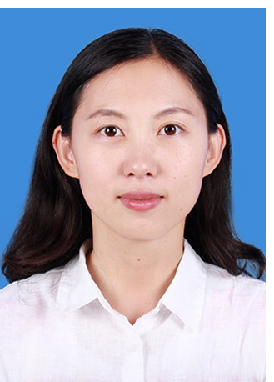}}]{Yuxing Han} received her B.S from Hong Kong University of Science and Technology, and Ph.D from UCLA, both in Electrical Engineering. \\
\indent Yuxing is currently a professor in school of engineering at South China Agriculture University, China. Her research area focuses on virtual reality, multimedia communication over challenging networks and big data analysis. She has authored many widely referenced papers and patents in related fields. Products deploying technologies that Dr. Han developed are currently widely used worldwide. The high quality video encoding/transcoding solution project Yuxing led won 2016 Frost \& Sullivan best practice in Enabling Technology Leadership award.
\end{IEEEbiography}

\end{document}